# Nanoscale Electromechanics of Paraelectric Materials with Mobile Charges: Size effects and Nonlinearity of Electromechanical Response of SrTiO$_3$ Films

*A.N. Morozovska[1], E.A. Eliseev[1,2], G.S. Svechnikov[1] and S.V. Kalinin[3a]*

[1] Institute of Semiconductor Physics, National Academy of Science of Ukraine,
41, pr. Nauki, 03028 Kiev, Ukraine

[2] Institute for Problems of Materials Science, National Academy of Science of Ukraine,
3, Krjijanovskogo, 03142 Kiev, Ukraine

[3] The Center for Nanophase Materials Sciences and Materials Sciences and Technology Division,
Oak Ridge National Laboratory, Oak Ridge, TN 37831

**Abstract**

Nanoscale enables a broad range of electromechanical coupling mechanisms that are forbidden or negligible in the materials. We conduct a theoretical study of the electromechanical response of thin paraelectric films with mobile vacancies (or ions) paradigmatic for capacitor-type measurements in X-ray scattering, piezoresponse force microscopy (PFM), and electrochemical strain microscopy (ESM). Using quantum paraelectric SrTiO$_3$ film as a model material with well known electromechanical, electronic and electrochemical properties, we evaluate the contributions of electrostriction, Maxwell stress, flexoelectric effect, deformation potential and compositional Vegard strains caused by mobile vacancies (or ions) and electrons to the electromechanical response. The local electromechanical response manifests strong size effects, the scale of which is determined by the ratio of the SrTiO$_3$ film thickness and PFM/ESM tip size to the carriers screening radius. Due to the strong dielectric nonlinearity effect inherent in quantum paraelectrics, the dependence of the SrTiO$_3$ film electromechanical response on applied voltage demonstrates a pronounced crossover from the linear to the quadratic law and then to the sub-linear law with a factor of 2/3 under the voltage increase. The temperature dependence of the electromechanical response as determined by the interplay between the dielectric susceptibility and the screening radius is non-monotonic and has a pronounced maxima, the position and width of which can be tuned by film thickness. This study provides a comparative framework for analysis of electromechanical coupling in the non-piezoelectric nanosystems.

---

[a]corresponding author: sergei2@ornl.gov

**I. Introduction**

The nanometer scale introduces a novel functionality in materials, with multiple examples including optical, magnetic, mechanical, and electronic transport properties [1, 2]. Many of these phenomena underpin multiple device, biological, or medical applications, and give rise to novel areas of scientific enquiry [3, 4, 5, 6]. In particular, nanoscale mechanical behaviors recognized as a vital component of nanoscience [7] are accessible as a result of evolutionary development from macroscopic testing to micron and nanoindentation [8], and subsequently to force-based Scanning Probe Microscopy (SPM) techniques [9]. Similarly, transport properties have been studied from macroscopic to molecular scales, both as the result of the development of new probing techniques and ever increasing demands of information technology [1]. The combination of a recognized need for understanding mechanical and electrical behavior at the nanoscale, and the evolution of measurement tools capable of addressing these properties on an ever decreasing length scale, has led to the present spectacular progress.

Of interest for nanoscale systems is the coupling between electrical, mechanical and transport phenomena including piezoresistive [10] and direct and converse electromechanical effects [11] in bulk materials and molecular systems. For the classical bulk piezo- and ferroelectric materials, the electromechanical coupling coefficients are typically small (~1-100 pm/V) [12, 13], thus requiring precise measurement even for macroscopic samples. Furthermore, for disordered materials such as (unpoled) polycrystalline ceramics and biological systems, the electromechanical properties described by antisymmetric tensors average to zero. These factors have limited quantitative and reproducible macroscopic studies of electromechanics to single crystals of materials such as quartz or ferroelectrics, recognized as important (microbalances, SAW, sonar, RF devices, ultrasonic imaging) [14, 15], but by now well studied class of systems.

The situation cannot be more different in the transition from the macroscopic to the nanometer scale. Nanoscale offers a broad array of novel electromechanical phenomena induced by symmetry breaking and low dimensionality that do not have macroscopic analogs. Examples include purely physical effects such as surface piezoelectricity and flexoelectricity [16, 17, 18, 19, 20, 21, 22, 23] as well as chemical and ionic processes in electrochemical [24, 25, 26, 27, 28, 29], molecular [30] and biological [31, 32] systems. These behaviors are enabled by larger strains that can be supported in nanoscale systems, as well as by the fact that local electroneutrality conditions are relaxed once system size becomes comparable or below the corresponding screening lengths (e.g. Debye lengths) [33]. This opens a new set of phenomena due to the electrostriction, Maxwell stress and deformation potential effects.

Despite the multitude of the electromechanical coupling mechanisms possible on the nanoscale, they are traditionally much less studied then optical, magnetic, and electronic effects. However, the situation has been changing rapidly in the last decade. Recent advances in scattering methods demonstrate that strain development in ultrathin films can be probed on the nanosecond scale [34, 35, 36, 37]. The strains can be also ascertained by the interferometric methods [38, 39, 40, 41]. The advances in aberration corrected scanning transmission electron microscopy now allow for direct mapping of atomic spacing, providing for observation of polarization, chemical effects, and strains on atomic (more specifically, single atomic column) level [42, 43, 44, 45, 46, 47]. Finally, techniques such as Piezoresponse Force Microscopy [48, 49] and Electrochemical Strain Microscopy [50, 51] allow studying time- and voltage electromechanical coupling on bare surfaces and in device structures locally, enabling direct observation of domain structures, polarization switching, and electrochemical reactivity.

Until recently, the nanoscale studies have been focused on material systems exhibiting macroscopic electromechanical responses such as piezoelectrics, ferroelectrics, and multiferroics. However, a large number of PFM/ESM studies of nominally non-polar materials such $SrTiO_3$ [52], CCTO [53] and LSMO [54] has been reported. While ionic motion could give rise to electromechanical signal (e.g. in electrochemical strain microscopy), the origins of contrast in materials such as STO and poor ionic conductors such as LSMO remain open. This motivates us to study all factors affecting electromechanical response in PFM on prototypical paraelectric material $SrTiO_3$ (STO), including the effects of concentration changes of the free charges due to diffusion and electromigration, flexoelectricity, electrostriction and Maxwell stresses. This analysis offers a model for other oxides, and also sets the context for future experimental studies.

## 2. Electromechanical response of paraelectrics

Experimental approaches for probing electromechanical coupling in nanoscale systems typically utilize local or global strain sensor coupled with the electrode material. In comparison, the current detection method directly uses detection of conduction (piezoresistance), displacement (ferroelectric) or Faradaic (ionics) currents. The comparative analysis of different detection principles suggests that strain detection offers significant advantages for nanoscale systems [55]. This approach is exemplified by PFM, in which the detection is performed using a small probe tip in contact with the free or electrode surface. In the former case, the tip acts both as an electrode and sensor, while in the latter case, it detects local deformations induced by the uniform field. This approach allows for spatial resolution and can be extended to time- and voltage spectroscopy methods [49]. The response in the capacitor structures can be tested by interferometric methods [40, 41]. These generally do not allow a spatial resolution, but offer a much higher z-resolution. Finally, laser vibrometers combine micron-scale lateral resolution with sub-nanometer sensitivity. Here, we study both top-electrode and tip-electrode geometries as shown in **Fig. 1**. Note, that even a very thin, flat capacitor is not equivalent to the SPM tip geometry, since in the latter case the lateral transport of electrons and ions is possible. Thus we will consider these cases separately: the *electromechanical response in planar geometry* is considered in Section 3, and the *local electromechanical response* (referred to as *PFM response*) is considered in Sections 4 and 5.

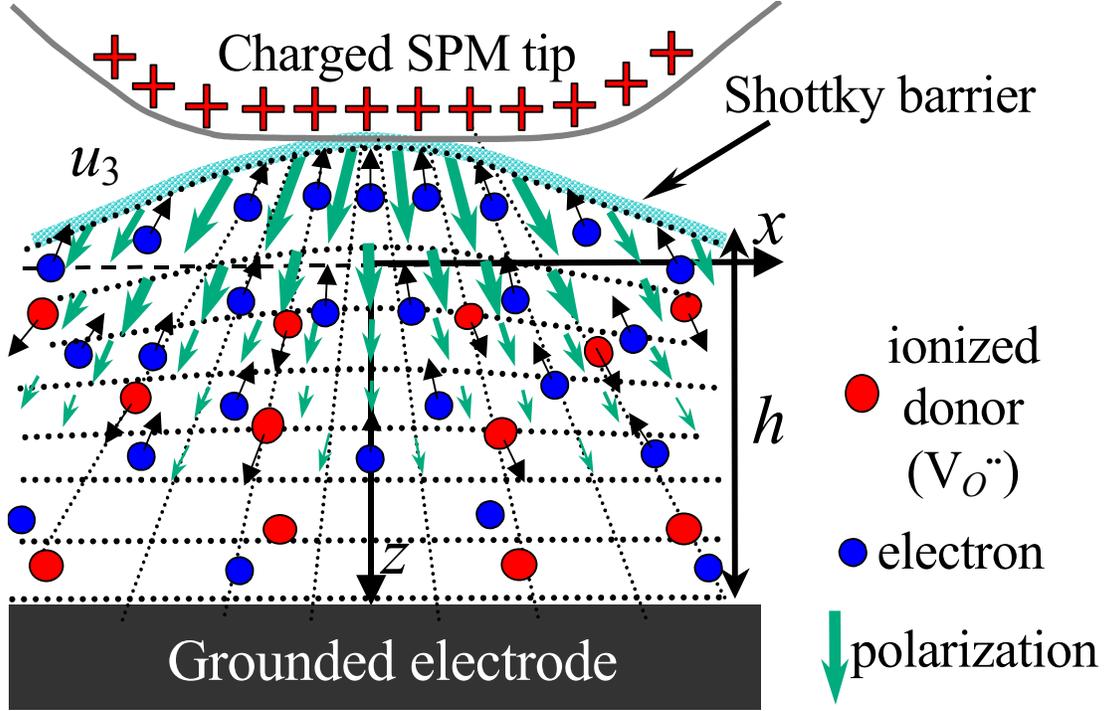

**Fig. 1.** Schematics of ESM/PFM measurements of the electromechanical response of the STO film placed between planar electrodes. The surface displacement counted from the fixed back interface is $u_3$; voltage $V_0$ is applied to the top electrode, $V_b$ is the built-in potential resulting from the Shottky barrier. Inhomogeneous polarization (arrows of different length and direction) induced by inhomogeneous electric field leads to the surface displacement via the flexoelectric and electrostriction effect. Compositional Vegard strains originate from the electron (blue balls) and ionized donors (red balls) electromigration due to the lattice dilatation.

**2.1. Generalized concentration-deformation free energy functional**

In order to model the electrical and elastic properties of ionic semiconductor in equilibrium, we derive the generalized expression for the free energy functional. Free energy for cubic symmetry paraelectrics including quantum corrections has the following form:

$$F = F_{ES} + F_{FLEXO} + F_{SE} + F_{CS} \tag{1a}$$

The first term in Eq.(1a) is the electrostatic energy of the quantum paraelectric with soft phonon mode, polarization gradient and space charge, that can be written down as

$$F_{ES} = \int_V d^3r \left( \begin{array}{c} \dfrac{\alpha(T)}{2} P_i P_i + \dfrac{\alpha_{11}}{4} P_i^2 P_i^2 + \dfrac{\alpha_{12}}{4} \sum_{i \neq j} P_i^2 P_j^2 + \dfrac{g_{ijkl}}{2} \left( \dfrac{\partial P_i}{\partial x_j} \dfrac{\partial P_k}{\partial x_l} \right) \\ - P_i E_i - \dfrac{\varepsilon_0 \varepsilon_b}{2} E_i E_i + e(N_d^+ - n) \varphi(\mathbf{r}) \end{array} \right) \quad (1b)$$

and is derived in **Appendix A.1**. Hereinafter $P_m(\mathbf{r})$ denotes electric polarization, $\varepsilon_b$ background permittivity and $\varepsilon_0 = 8.85 \times 10^{-12}$ F/m the dielectric permittivity of vacuum. For quantum paraelectrics such as SrTiO$_3$, KTaO$_3$, and EuTiO$_3$ in cubic phase [56, 57, 58], the expansion coefficient $\alpha$ is temperature dependent in accordance with Barrett Law, $\alpha(T) = \alpha_T \left( \coth(T_q/(2T)) T_q/2 - T_0 \right)$. The higher order coefficients $\alpha_{ij}$, including the 4$^{\text{th}}$ order terms in polarization, are relevant for cubic materials with soft mode nonlinearity. These and other quantities, including gradient coefficient $g_{ijkl}$, are listed and defined in the **Table 1.** Note that for high temperatures Eq.(1b) is applicable to all cubic perovskites, e.g. for ferroelectrics in paraelectric phase above the Curie temperature. Film-substrate misfit strains can be accounted through the renormalization of the free energy coefficients $\alpha_i(T)$ and $\alpha_{ij}$ [59]. However below we consider only matched substrates (film-substrate misfit strain is negligibly small), or thick slabs (where misfit strain is relaxed [60]) and defer the discussion of the effect of strain-induced ferroelectric phase transition [61, 62] to future studies.

In Eq. (1b), $E_k(\mathbf{r}) = -\partial \varphi(\mathbf{r})/\partial x_k$ denotes the electric field, $\varphi(\mathbf{r})$ the electric potential, $e(N_d^+ - n)$ the space charge density, $e = 1.6 \times 10^{-19}$ C the electron charge, $n(\mathbf{r})$ the concentration of the electrons in the conduction band and $N_d^+(\mathbf{r})$ the concentration of ionized defects (e.g. oxygen vacancies) which could be mobile. The latter term in Eq.(1b) is the electrostatic energy of free charges with density $e(N_d^+ - n)$ in the electric field with potential $\varphi$ (e.g. Ref.[63]).

The second term in Eq.(1a) is the flexoelectric effect contribution:

$$F_{FLEXO} = \int_V d^3r \frac{f_{ijkm}}{2} \left( u_{ij} \frac{\partial P_m}{\partial x_k} - P_m \frac{\partial u_{ij}}{\partial x_k} \right). \quad (1c)$$

Hereinafter $u_{kl}(\mathbf{r})$ denotes the elastic strain and $f_{ijkm}$ the flexoelectric tensor. The third term in Eq.(1a) is the contribution of electrostriction coupling and elastic energy:

$$F_{SE} = \int_V d^3r \left( u_{ij} \left( q_{ijkl} P_k P_l + T_{ij}(\mathbf{D}, \mathbf{E}) \right) + \frac{c_{ijkl}}{2} u_{ij} u_{kl} \right) \quad (1d)$$

Here $q_{ijkl}$ denotes the electrostriction stress tensor, $c_{ijkl}$ the elastic stiffness tensor and $T_{ij}(\mathbf{D},\mathbf{E}) = \left( D_i E_j - \frac{\delta_{ij}}{2} D_m E_m \right) \approx \varepsilon_0 \varepsilon \left( E_i E_j - \frac{\delta_{ij}}{2} E_m E_m \right)$ the Maxwell stress tensor [64, 65]. For polarized media of cubic symmetry with isotropic dielectric permittivity tensor $\varepsilon_{ij} = \varepsilon \delta_{ij}$ the electric displacement is given as $D_k = P_k + \varepsilon_0 E_k$.

The last term in Eq. (1a) represents the contributions of concentration-strain coupling and configuration entropy of the charged species:

$$F_{CS} = \int_V d^3r \left( \left( \Xi_{ij}(n - n_e) + \beta_{ij}(N_d^+ - N_{de}^+) \right) u_{ij} + k_B T \, S(n_e, n) + k_B T \, S(N_{de}^+, N_d^+) \right) \quad (1e)$$

In the Boltzmann-Planck-Nernst approximation, the configuration entropy function is given as $S(x,y) = y \ln(y/x) - y$; $k_B = 1.3807 \times 10^{-23}$ J/K, where $T$ is the absolute temperature. Equation (1e) includes **electrochemical** concentration-deformation energy, $\left( \Xi_{ij}(n - n_e) + \beta_{ij}(N_d^+ - N_{de}^+) \right) u_{ij}$, which is determined by the convolution of the tensorial deformation potential tensor $\Xi_{ij}$ and Vegard expansion tensor $\beta_{ij}$ with elastic strain tensor $u_{jk}(\mathbf{r})$. In the absence of external potential and strains, the **equilibrium concentrations** of the electrons in the conduction band and ionized defects are represented as $n_e$ and $N_{de}^+$ respectively. Typical values of $n_e$ and $N_{de}^+$ for SrTiO$_3$ with mobile oxygen vacancies are listed in the **Table 1.**

The functional in Eq.(1a) is the Helmholtz free energy with the strain field as its independent variable [66, 67, 68, 69, 70]. Other forms of free energy functional (e.g. Gibbs form) could be deduced from Eqs.(1) using corresponding Legendre transformations.

Finally, here we neglect the surface energy contribution in free energy (1), in particular the surface piezoelectric effect, originated from the inversion center absence in the immediate vicinity of surface [71, 72, 73]. Despite the fact that the surface piezoelectric effect should exist as required by symmetry theory, the magnitudes of its coefficients are still a challenge and are likely to be small for an ideal surface (also note that surface behavior is likely to be dominated by the ionic screening conditions, as analyzed by Stephenson et al [74]).

**Table 1.** Material parameters for SrTiO$_3$.

| Parameter | Unit | Quantum paraelectric SrTiO$_3$ | Notes and Refs |
|---|---|---|---|
| Background permittivity $\varepsilon_b$ | dimensionless | 3 – 43 | * [75] |
| Full permittivity $\varepsilon = \left(\varepsilon_b + \dfrac{1}{\alpha\varepsilon_0}\right)$ | dimensionless | 297-300 | * at RT |
| Soft mode related permittivity $\varepsilon_{QP} = (\alpha\varepsilon_0)^{-1}$ | dimensionless | 257-207 | * at RT |
| LGD-expansion coefficient $\alpha_T$ | 10$^6$ m/(F K) | 1.66 | * [56, 57] |
| Curie temperature $T_0$ | K | 36 | * [56, 57] |
| Quantum vibration temperature $T_q$ | K | 100 | * [56, 57] |
| LGD-gradient coefficient $g$ | 10$^{-10}$ V·m$^3$/C | 1 – 10 | [76] |
| LGD-expansion coefficients $\alpha_{ijkl}$ for elastically **free** system | 10$^9$ m$^5$/(C$^2$F) | $\alpha_{11}$=8.1 $\alpha_{12}$=2.4 | * [56, 57] |
| LGD-expansion coefficients $\alpha_{ijkl}$ for elastically **clamped** system | 10$^9$ m$^5$/(C$^2$F) | $\alpha_{11}$=9.6 $\alpha_{12}$=3.2 | * [56, 57] |
| Electrostriction strain coefficients $Q_{ijkl}$ | m$^4$/C$^2$ | $Q_{11}$=0.051 $Q_{12}$= –0.016 $Q_{44}$=0.020 | * [77] |
| Electrostriction stress coefficients $q_{ijkl}=-c_{ijnm}Q_{nmkl}$ (Vogt notation) | 10$^9$mJ/C$^2$ | $q_{11}$=-13.7 $q_{12}$= 1.6 $q_{44}$=-2.5 | * recalculated from $Q_{ijkl}$ and $c_{ijkl}$ |
| Flexoelectric tensor $\gamma_{ijkl} = f_{ijkl}/\alpha$ at room temperature (300 K) (Vogt notation for cubic symmetry) | 10$^{-9}$C/m | $\gamma_{11}$= – 9 $\gamma_{12}$= 4 $\gamma_{44}$= 3 | measured by [78] only once |
| Elastic stiffness $c_{ij}$ | 10$^{11}$ N/m$^2$ | $c_{11}$=3.36 $c_{12}$=1.07 $c_{44}$=1.27 | * [77] |
| Elastic compliances $s_{ij}$ | 10$^{-12}$ m$^2$/N | $s_{11}$=3.52 $s_{12}$= –0.85 $s_{44}$=7.87 | * recalculated from stiffness $c_{ijkl}$ |
| Vegard expansion tensor $\beta_{ij}$ (for cubic symmetry $\beta_{ij} = \beta\delta_{ij}$) | eV | ~1 | order of magnitude taken from [79, 80, 81] |
| Tensorial deformation potential $\Xi_{ij}$ (for cubic symmetry $\Xi_{ij} = \Xi\delta_{ij}$) | eV | 2.87 | estimated as 2$E_F$/3 for $E_F$=4.3 eV * [82, 83] |
| Equilibrium concentration of the free electrons $n_e$ and mobile ions $N_{de}^+$ (oxygen vacancies) | m$^{-3}$ | 10$^{23}$ – 10$^{26}$ | * [84, 85] at RT, depending on the oxygen |

* means that the parameter value is well known, RT – room temperature

## 2.2. Thermodynamic equilibrium: equations of state with boundary conditions

A variation of the free energy functional Eq.(1) with respect to its independent variables (potential φ, polarization P and strain tensor $u_{ij}$) allows us to derive corresponding equations of state. These, along with the corresponding boundary conditions, are discussed below.

1) Equation of state for polarization follows from the minimization of the free energy (1) $\delta F/\delta P_i = 0$ (here, δ represents the variation derivative [86]) as

$$\left(\alpha\delta_{ij} + 2u_{mn}q_{mnij}\right)P_j + \alpha_{ijkl}P_jP_kP_l - g_{ijkl}\frac{\partial^2 P_k}{\partial x_j \partial x_l} = f_{mnli}\frac{\partial u_{mn}}{\partial x_l} + E_i - \left(u_{ij}E_j - \frac{u_{kk}}{2}E_i\right) \quad (2a)$$

The Maxwell stress term $\left(u_{ij}E_j - \frac{u_{kk}}{2}E_i\right)$ is typically very small in comparison with $E_i$, since the absolute values of strain are always very small in comparison with unity, $|u_{ij}| \ll 1$. If nonlinearity and gradient effects are not taken into consideration, then Eq.(2a) yields $P_i \approx \left(\alpha\delta_{ij} + 2u_{mn}q_{mnij}\right)^{-1}\left(f_{mnkj}(\partial u_{mn}/\partial x_k) + E_j\right)$ ($\delta_{ij}$ is Kroneker delta).

Allowing for the gradient term contribution to free energy (1b), the natural boundary conditions for Eq.(2a) are obtained after the minimization of free energy:

$$g_{ijkl}n_j\frac{\partial P_k}{\partial x_l}\bigg|_S = 0 \quad (2b)$$

Note that this condition is consistent with the quasi-homogeneous distribution of polarization, since we do not consider the surface energy contribution to free energy, which leads to the intrinsic distribution of spontaneous polarization in nanosized ferroelectrics [87].

2) Variation $\frac{\delta F}{\delta \varphi} = \frac{\delta F_{ES}}{\delta \varphi} = 0$ leads to the Poisson-type equation for electrostatic potential $\varphi(\mathbf{r})$:

$$\varepsilon_0\varepsilon_b\frac{\partial^2\varphi}{\partial x_i \partial x_i} = -e(N_d^+ - n) + \frac{\partial P_k}{\partial x_k}. \quad (3a)$$

Boundary conditions to Eq.(3a) have the form:

$$\varphi(x_1, x_2, 0) = V_0(r) + V_b, \quad \varphi(x_1, x_2, h) = 0. \quad (3b)$$

where $V_0(r)$ denotes the radially symmetric electrostatic potential distribution produced by the PFM tip at the sample surface z=0, $r = \sqrt{x_1^2 + x_2^2}$ and h is the film thickness. $V_b$ represents the constant built-in potential [88, 89], which originate from surface dipole layers, e.g. due to the Shottky barrier at the tip (or electrode) – surface junction. The planar electrode at $z=h$ is regarded ohmic, i.e. the potential is continuous here.

Concentrations of the electrons and ionized donors in the Boltzmann-Planck-Nernst approximation are

$$n(\mathbf{r}) = n_0 \exp\left(\frac{-\Xi_{ij} u_{ij}(\mathbf{r}) + e\varphi(\mathbf{r})}{k_B T}\right), \quad N_d^+(\mathbf{r}) = N_{d0}^+ \exp\left(\frac{-\beta_{jk} u_{jk}(\mathbf{r}) - e\varphi(\mathbf{r})}{k_B T}\right). \tag{4}$$

Note, that no principal constrains exist on the values of $n_0$ or $N_{d0}^+$ for the case of *ion* or *electron conducting* interfaces (**CI**); however, they are dependent on the electrochemical potential of the material and electrodes. Here, we choose the special case of boundary conditions $n_0 \equiv n_e$ and/or $N_{d0}^+ \equiv N_{de}^+$ for electron/ion CI respectively. The condition $N_{d0}^+ = n_0$ holds for a semi-infinite paraelectric, since the electric field should vanish in the material depth. For the films of small thickness, electro-neutrality of the whole system "film + electrodes" can be maintained by the free charges accumulated at the planar electrodes and $N_{d0}^+ \neq n_0$ is possible in thin films due to the carriers injection/divergence.

For the case of *blocking interfaces* (**BI**) constrains, the $n_0$ and $N_{d0}^+$ values can be derived from appropriate conservation laws. When both interfaces $z = 0$ and $h$ are *ion-blocking*, the total amount of donors remain constant $\int_V N_d^+(\mathbf{r}) d^3r = V N_{de}^+ = const$ in the material of volume $V$, and

$$N_{d0}^+ = N_{de}^+ V \bigg/ \int_V \exp\left(\frac{-\beta_{jk} u_{jk}(\mathbf{r}) - e\varphi(\mathbf{r})}{k_B T}\right) d^3r. \tag{5a}$$

When both interfaces are *electron-blocking*, the total amount of electrons are constant $\int_V n(\mathbf{r}) d^3r = V n_e = const$ and

$$n_0 = n_e V \bigg/ \int_V \exp\left(\frac{-\beta_{jk} u_{jk}(\mathbf{r}) + e\varphi(\mathbf{r})}{k_B T}\right) d^3r. \tag{5b}$$

In the case when even one of the interfaces are *ion(electron)-conducting,* no such constrains are present in the thermodynamic equilibrium, while kinetic processes (current flow) essentially depend on the blocking/conducting conditions at both interfaces.

3) Variation of the free energy (1) on the strain tensor gives the stress tensor, $\sigma_{ij}(\mathbf{r}) = \delta F / \delta u_{ij}$, as:

$$\sigma_{ij}(\mathbf{r}) = \begin{pmatrix} c_{ijkl} u_{kl}(\mathbf{r}) + \left(\Xi_{ij}(n(\mathbf{r}) - n_e) + \beta_{ij}(N_d^+(\mathbf{r}) - N_{de}^+)\right) + f_{ijkl} \dfrac{\partial P_l}{\partial x_k} \\ + q_{ijkl} P_k P_l + \left(\delta_{ik}\delta_{jl} - \dfrac{\delta_{ij}\delta_{kl}}{2}\right)(P_k + \varepsilon_0 E_k) E_l \end{pmatrix}. \quad (6)$$

As the strain tensor is given by $u_{ij} = \dfrac{1}{2}\left(\dfrac{\partial u_i}{\partial x_j} + \dfrac{\partial u_j}{\partial x_i}\right)$, the Lame-type equation for mechanical displacement $u_i$ can be obtained from the equation of mechanical equilibrium $\partial \sigma_{ij}(\mathbf{r})/\partial x_i = 0$, where the stress tensor $\sigma_{ij}(\mathbf{r})$ is given by Eq.(6), namely:

$$c_{ijkl} \dfrac{\partial^2 u_k}{\partial x_j \partial x_l} = -\dfrac{\partial}{\partial x_j}\begin{pmatrix} \left(\Xi_{ij}(n(\mathbf{r}) - n_e) + \beta_{ij}(N_d^+(\mathbf{r}) - N_{de}^+)\right) + f_{ijkl} \dfrac{\partial P_l}{\partial x_k} \\ + q_{ijkl} P_k P_l + \left(\delta_{ik}\delta_{jl} - \dfrac{\delta_{ij}\delta_{kl}}{2}\right)(P_k + \varepsilon_0 E_k) E_l \end{pmatrix} \quad (7)$$

Mechanical boundary conditions [90] corresponding to the PFM/ESM experiments [49, 50, 55] are defined on a mechanically free interface, $z = 0$, where the normal stress is absent (more specifically, tip-surface forces are small), and on substrate interface $z = h$, where the displacement $u_i$ is zero for a thick "rigid" substrate or continuous for a "soft" thin substrate:

$$\sigma_{3i}(x_1, x_2, z = 0) = 0, \quad (8a)$$

$$u_i(x_1, x_2, z = h) = 0 \quad \text{or} \quad u_i(x_1, x_2, z = h - 0) = u_i(x_1, x_2, z = h + 0). \quad (8b)$$

**2.3. The electromechanical response of the film surface**

The electromechanical response of the film surface at the point $x_3=0$, i.e. surface displacement at the tip-surface junction detected by SPM electronics, for elastically isotropic semi-space, can be calculated as:

$$u_i(x_1, x_2, z) = u_{CS} + u_{FLEXO} + u_{ES} + u_{MT}. \quad (9a)$$

For cubic symmetry, the four contributions in Eq.(9a) are concentration-strain, flexoelectric, electrostriction, and Maxwell strain respectively. Their explicit forms are following.

***Concentration-strain*** contribution including compositional Vegard strain and deformation potential:

$$u_{CS} = -\iiint\limits_{0<\xi_3<h} \frac{\partial G_{ij}^S(x_1-\xi_1, x_2-\xi_2, z, \xi_3)}{\partial \xi_m} \left(\Xi_{mj}(n(\xi)-n_e) + \beta_{mj}(N_d^+(\xi)-N_{de}^+)\right) d^3\xi, \quad (9b)$$

Here, $G_{ij}^S$ denotes the tensorial elastic Green function corresponding to a semi-space (the case $h \to \infty$) as listed in e.g. Ref.[91]) or to a thin film placed on a rigid or matched substrate (derived in Refs.[92, 93]).

***Flexoelectric*** contribution:

$$u_{FLEXO} = -\iiint\limits_{0<\xi_3<h} \frac{\partial G_{ij}^S(x_1-\xi_1, x_2-\xi_2, z, \xi_3)}{\partial \xi_m} f_{mjkl} \frac{\partial P_l}{\partial \xi_k} d^3\xi \quad (9c)$$

***Electrostriction*** contribution:

$$u_{ES} = -\iiint\limits_{0<\xi_3<h} \frac{\partial G_{ij}^S(x_1-\xi_1, x_2-\xi_2, z, \xi_3)}{\partial \xi_m} q_{mjkl} P_k P_l d^3\xi, \quad (9d)$$

***Maxwell stress*** contribution:

$$u_{MT} = -\iiint\limits_{0<\xi_3<h} \frac{\partial G_{ij}^S(x_1-\xi_1, x_2-\xi_2, z, \xi_3)}{\partial \xi_m} \left(\delta_{mk}\delta_{jl} - \frac{\delta_{mj}\delta_{kl}}{2}\right)(P_k + \varepsilon_0 E_k)E_l d^3\xi. \quad (9e)$$

Explicit form of Eqs. (9b)-(9e) is listed in Appendix B for materials with cubic symmetry.

## 2.4. Electromechanical response in decoupling approximation

Equations (9a-e), (3a-b) and (2) form the nonlinear coupled system. Below, we introduce a decoupling approximation in which we consider the flexoelectric effect. The electrostriction contribution is small enough to be disregarded in the Poisson equation. The accuracy of the approximation is discussed in the next subsection. The decoupling approximation allows the individual terms in Eq.(9a) to be evaluated as following:

**Step 1.** Determination of the electric potential from the Eq.(3a) without strain terms:

$$\left(\varepsilon_0 \varepsilon_b + \frac{1}{\alpha}\right)\Delta\varphi = -e(N_d^+ - n) \quad (10a)$$

Where

$$N_d^+(\mathbf{r}) = N_{d0}^+ \exp\left(\frac{-e\varphi(\mathbf{r})}{k_B T}\right) \quad \text{and} \quad n(\mathbf{r}) = n_0 \exp\left(\frac{e\varphi(\mathbf{r})}{k_B T}\right). \quad (10b)$$

In linear Debye approximation, mainly used hereinafter, i.e. regarding that $|e\varphi/k_B T| \ll 1$ and expanding the exponents in Eq.(10b) as $\exp(x) \approx 1+x$ we get

$$N_d^+ \approx N_{d0}^+\left(1 - \frac{e\varphi}{k_B T}\right) \text{ and } n \approx n_0\left(1 + \frac{e\varphi}{k_B T}\right), \tag{10c}$$

the solution of the Poisson equation (10a) with boundary conditions (5) was determined as:

$$\varphi(r,z) = \begin{pmatrix} e\dfrac{(N_{d0}^+ - n_0)R_d^2}{\varepsilon_0 \varepsilon}\left(1 - \dfrac{\exp((h-z)/R_d) + \exp(z/R_d)}{1 + \exp(h/R_d)}\right) + \\ V_b\left(\dfrac{\exp(-z/R_d) - \exp(-(2h-z)/R_d)}{1 - \exp(-2h/R_d)}\right) + \int_0^\infty dk\, k J_0(kr) \cdot \widetilde{\varphi}_V(k,z) \end{pmatrix} \tag{11a}$$

$$\widetilde{\varphi}_V(k,z) = \widetilde{V}_0(k)\frac{\exp(-K(k)z) - \exp(-K(k)(2h-z))}{1 - \exp(-2K(k)h)} \tag{11b}$$

Where $K(k) = \sqrt{k^2 + R_d^{-2}}$, since above the temperature of the structural phase transition (~105 K) the dielectric permittivity is isotropic: $\varepsilon = \left(\varepsilon_b + \dfrac{1}{\alpha \varepsilon_0}\right)$. The Debye screening radius is introduced as $R_D = \sqrt{\dfrac{\varepsilon_0 \varepsilon k_B T}{e^2(n_0 + N_{d0}^+)}}$.

**Step 2.** Substitution of the polarization from Eq.(2), disregarding the strain terms, $P_i \approx -(\alpha^{-1})\partial\varphi/\partial x_i$, in Eq.(9a) leads to

$$u_i(x_1,x_2,z) = -\iiint_{0<\xi_3<h} \frac{\partial G_{ij}^S(x_1-\xi_1, x_2-\xi_2, z, \xi_3)}{\partial \xi_m}\begin{pmatrix}\Xi_{mj}(n(\xi)-n_e) + \beta_{mj}(N_d^+(\xi) - N_{de}^+) \\ + \dfrac{q_{ijkl}^{MT}}{\alpha^2}\dfrac{\partial\varphi}{\partial \xi_k}\dfrac{\partial\varphi}{\partial \xi_l} - \dfrac{f_{mjkl}}{\alpha}\dfrac{\partial^2 \varphi}{\partial \xi_k \partial \xi_l}\end{pmatrix} d^3\xi \tag{12a}$$

Where the electrostriction tensor is renormalized by the Maxwell stress as

$$q_{ijkl}^{MT} = q_{ijkl} + \frac{\varepsilon}{\varepsilon_0(\varepsilon-1)^2}\left(\delta_{ik}\delta_{jl} - \frac{\delta_{ij}\delta_{kl}}{2}\right) \tag{12b}$$

In fact, Eq.(12b) shows that the electrostriction term acts as the "electrostatic force" effect.

For cubic symmetry the Eq.(12a) contains three contributions: ***concentration-strain, electrostriction*** renormalized by the ***Maxwell stress*** in accordance with Eq.(12b) and the ***flexoelectric*** contribution (see **Appendix A**). Interestingly, the contribution of Maxwell stress to (12b) is estimated to be negligibly small for SrTiO$_3$, since $q_{ijkl} \sim 10^9 - 10^{10}$ mJ/C$^2$ (see **Table 1**), while $\frac{\varepsilon}{\varepsilon_0(\varepsilon-1)^2} \cong \frac{1}{\varepsilon_0 \varepsilon} = 3.8 \times 10^7$ mJ/C² at room temperature, so their ratio $2q_{11}\varepsilon_0(\varepsilon-1)^2/\varepsilon \approx 50$.

It is obvious that for the case $2q_{11}\varepsilon_0(\varepsilon-1)^2/\varepsilon \leq 1$ Maxwell's stress would be comparable with electrostriction for the considered geometry of film on substrate electrode. For instance, estimations for fused silica, SiO$_2$, gives $2q_{11}\varepsilon_0(\varepsilon-1)^2/\varepsilon \approx 0.28$, and $2q_{11}\varepsilon_0(\varepsilon-1)^2/\varepsilon \approx 4$ for MgO (electrostriction coefficients are taken from Ref.[94]).

Equations (12a-b), (11a-b) and (10b) form the linear decoupled system that can be solved analytically. Within the decoupling approximation, the concentration-strain and flexoelectric contributions are linear with respect to the electric field (and hence applied voltage) and thus they mimic "piezoelectric like", while electrostriction and Maxwell stress contribution are quadratic with respect to the applied voltage.

**2.5. Accuracy of the decoupling approximation. Polarization nonlinearity impact**

To estimate the accuracy of the decoupling approximation we studied numerically and semi-analytically the following 1D-model:

(a) All variables are only z-dependent in the coupled system (2), (3), (9).

(b) Debye approximation Eq. (10c) is used for electrostatic potential.

(c) The assumption $N_{d0}^+ = n_0$ is introduced, i.e. formation of each immobile ion produces an electron.

Within the coupled model we minimized the free energy (1) numerically and analytically by the direct variational method [72] and then derived an approximate analytical solution for the electric field $E_3(z)$, polarization $P_3(z)$ and strain $u_{33}(z)$ distributions (see **Appendix A.3**). The approximate analytical dependence of the displacement (9) on applied voltage has the form:

$$u_3^{coupled}(0) = \begin{cases} \left( n_0 \frac{Ue}{k_B T} \frac{\Xi - \beta}{c_{11}} \frac{h}{2} - \frac{f_{11}}{c_{11}} \frac{U}{\tilde{\alpha}} \frac{h^3}{2\tilde{R}_d^2(12\tilde{g}\kappa + h^2)} + \frac{q_{11}}{c_{11}} h^{1/3} \left(\frac{U}{\tilde{\alpha}_{11}}\right)^{2/3} \right), & h \leq R_d, \\ \left( n_0 \frac{Ue}{k_B T} \frac{\Xi - \beta}{c_{11}} \frac{\tilde{R}_d^2 + \tilde{R}_d\sqrt{\kappa \tilde{g}}}{L_c} - \frac{f_{11}}{c_{11}} \frac{U}{\tilde{\alpha} L_c} + \frac{U^2}{\tilde{\alpha}^2} \frac{q_{11}}{c_{11}} \left(\frac{1}{2L_c} + \frac{\tilde{R}_d\sqrt{\kappa \tilde{g}}}{2L_c^3}\right) \right), & h \gg R_d \end{cases}$$

(13a)

Here we introduced the **total potential drop** $U = V_0 + V_b$, coefficient $\tilde{\alpha}_{11} = \alpha_{11} - 2q_{11}^2/c_{11}$ renormalized by electrostriction, inverse susceptibility $\tilde{\alpha}$ renormalized by paraelectric nonlinearity $\sim \langle P \rangle^2$, and permittivity $\kappa = \varepsilon_0 \varepsilon_b$, renormalized Debye screening radius $\tilde{R}_d$ and gradient term $\tilde{g}$ as:

$$\tilde{\alpha} = \alpha + 3\tilde{\alpha}_{11}\langle P \rangle^2, \quad \tilde{R}_d^2 = \frac{k_B T}{2N_{d0}^+ e^2}\left(\varepsilon_0 \varepsilon_b + \frac{1}{\tilde{\alpha}}\right), \quad \tilde{g} \equiv \left(g_{33} - \frac{f_{33}^2}{c_{11}}\right). \tag{13b}$$

The characteristic length $L_c = \sqrt{\tilde{R}_d^2 + 2\tilde{R}_d\sqrt{\kappa\tilde{g}} + \frac{\tilde{g}}{\tilde{\alpha}}}$ includes the combination of the screening radius $\tilde{R}_d$ and correlation length of paraelectric $l_c = \sqrt{\tilde{g}/\tilde{\alpha}}$ (and we regard it as the screening-correlation length). Estimations for STO parameters gives $\sqrt{\kappa\tilde{g}} \sim 0.4$ nm, $\tilde{R}_d \leq 10$ nm and $\tilde{g}/\tilde{\alpha} \sim 1$ nm for $U$=0.

Note that the difference between $\tilde{\alpha}$ and $\alpha$ depends on the applied voltage, since average polarization $\langle P_3 \rangle = \frac{1}{h}\int_0^h P_3 dz$ depends on the potential $U$ and film thickness $h$ via the cubic equation $\left(\alpha + \tilde{\alpha}_{11}\langle P_3 \rangle^2\right)\langle P_3 \rangle = U/h$, where the right-hand side is the average electric field, $\frac{1}{h}\int_0^h E_3 dz \equiv \frac{U}{h}$, independent on the Debye screening $R_d$. The solution of the cubic equation has the form:

$$\langle P_3 \rangle = p(h,U)\left(1 - \frac{\alpha}{3\tilde{\alpha}_{11}p(h,U)^2}\right) \approx \begin{cases} \dfrac{U}{\alpha h}, & \dfrac{U}{h} \ll E_c, \\ \left(\dfrac{U}{\tilde{\alpha}_{11}h}\right)^{1/3}, & \dfrac{U}{h} \gg E_c. \end{cases} \tag{14a}$$

Where the amplitude $p(h,U)$ and characteristic field $E_c$ are:

$$p(h,U) = \left(\frac{U}{2\tilde{\alpha}_{11}h} + \sqrt{\left(\frac{\alpha}{3\tilde{\alpha}_{11}}\right)^3 + \left(\frac{U}{2\tilde{\alpha}_{11}h}\right)^2}\right)^{1/3}, \quad E_c = 2\tilde{\alpha}_{11}\left(\frac{\alpha}{3\tilde{\alpha}_{11}}\right)^{3/2} \tag{14b}$$

The characteristic field $E_c$ mimics the thermodynamic coercive field of ferroelectrics with negative α, $E_c^{FE} = 2\tilde{\alpha}_{11}\left(\dfrac{-\alpha}{3\tilde{\alpha}_{11}}\right)^{3/2}$. The characteristic field is $E_c = 3.9 \times 10^7$ V/m for STO parameters at room temperature. The field $E_c$ decreases with temperature, since α falls with temperature in accordance with Barrett Law. Two limiting cases of Eq.(14a) correspond to a low and a high effective field $E^* \sim U/h$.

In the decoupling approximation (i.e. assuming $P_i \approx -(\alpha^{-1})\partial\varphi/\partial x_i$ as described in subsection 2.4) the surface displacement has the form:

$$u_3^{decoupled}(0) = \begin{cases} n_0 \dfrac{eU}{k_B T}\dfrac{(\Xi-\beta)}{c_{11}}\dfrac{h}{2} - \dfrac{f_{11}}{c_{11}}\dfrac{U}{\alpha}\dfrac{h}{2R_d^2} + \dfrac{U^2}{\alpha^2}\dfrac{q_{11}}{c_{11}}\dfrac{1}{h}, & h \leq R_d, \\ n_0 \dfrac{Ue}{k_B T}\dfrac{\Xi-\beta}{c_{11}}R_d - \dfrac{f_{11}}{c_{11}}\dfrac{U}{\alpha}\dfrac{1}{R_d} + \dfrac{U^2}{\alpha^2}\dfrac{q_{11}}{c_{11}}\dfrac{1}{2R_d}, & h \gg R_d. \end{cases} \quad (15)$$

As anticipated, the coupled solution Eq.(13a) yields the decoupled solution Eq.(15) in the limiting case where $\tilde{g} \to 0$ and $\tilde{\alpha}_{11} \to 0$.

Note that the difference between the displacement calculated in the coupled problem (Eq.(13)-(14)) and decoupled approximation (Eq.(15)) are determined by the thickness dependence (especially important at small thicknesses $h \leq R_d$) and coefficients $\tilde{R}_d$ and $\tilde{\alpha}$ renormalization. In particular, in the coupled problem, the striction contribution scales as $h^{1/3}(U/\tilde{\alpha}_{11})^{2/3}$ as the film thickness decreases, compared to the behavior expected in linear decoupled model, where it increase as $U^2/(\alpha^2 h)$ with thickness decrease. The origin of the electrostriction contribution divergence in the pure decoupling approximation at small thicknesses is the divergence of the electric field, since electrostriction strain per se is proportional to the squire of the electric field, that is in turn proportional to the film thickness square: $u_{ii} \sim E_i^2 \sim (U/h)^2$. Integration of the electrostriction strain yields $u_3(0) \sim \int_0^h dz\, u_{ii} \sim h u_{ii} \sim 1/h$. The rapid increase of the electrostriction contribution starts at the critical thickness $h_{cr} \sim R_d$.

From Eqs.(13)-(15) it is obvious that the electrostriction contribution dominates at high applied voltages and/or small film thicknesses. Correspondingly, in this region the decoupling approximation accuracy is not satisfactory. Accuracy of the decoupling approximation should be high enough for thick films with $h >> R_d$.

Comparison between numerical simulations of the coupled problem and analytical expressions obtained in decoupling approximation is shown in **Fig.2.**

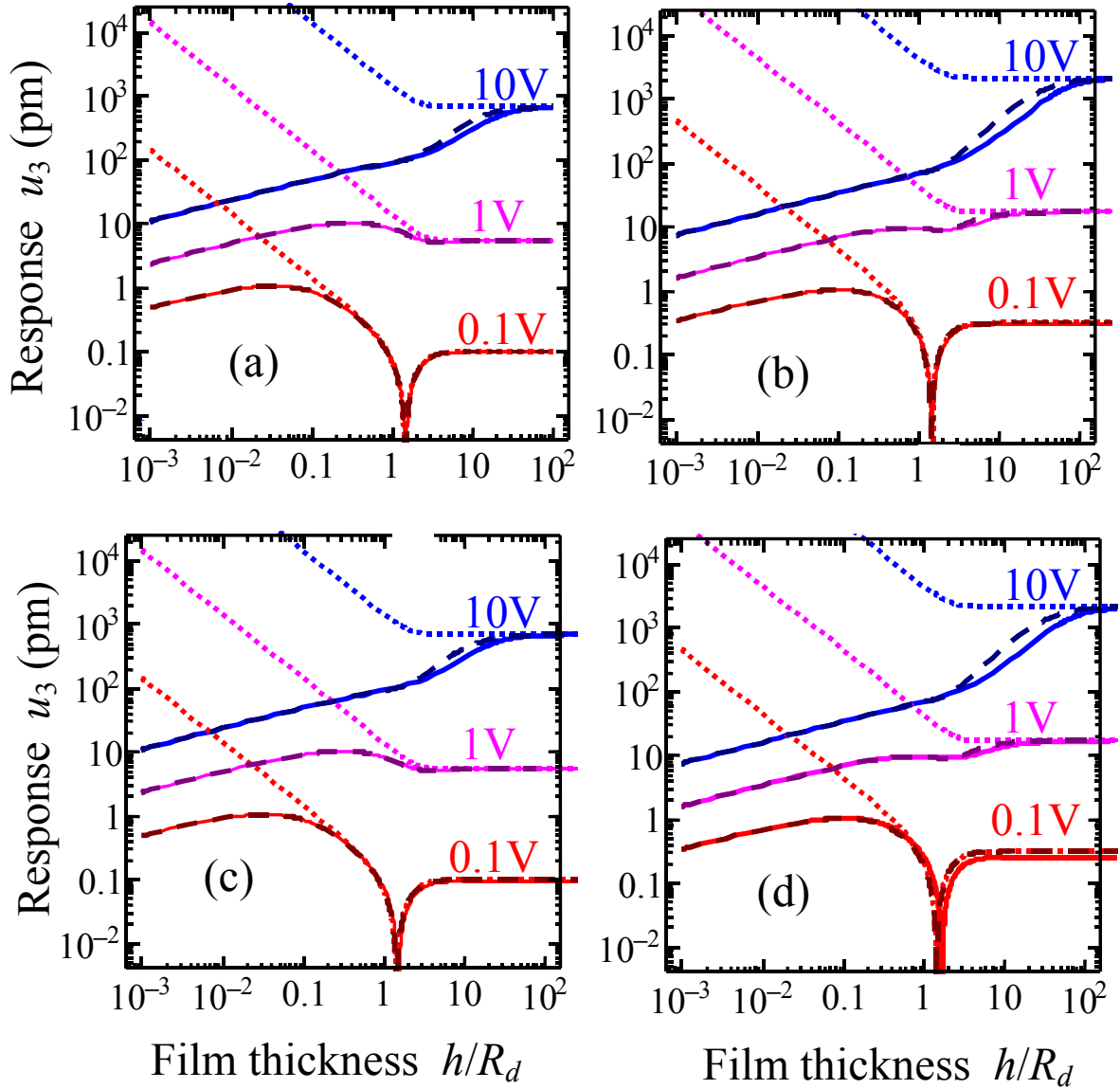

**Fig.2.** Thickness dependence of one-dimensional electromechanical response amplitude calculated for STO film and different total potentials $U$=0.1 V, 1 V, 10 V (figures near the curves); $n_0=N_{d0}=10^{24}$ m$^{-3}$ (a, b); $n_0=N_{d0}=10^{25}$ m$^{-3}$ (c, d); gradient coefficient g = $10^{-10}$ V·m$^3$/C (a, c); g = $10^{-9}$ V·m$^3$/C (a, c). Dotted curves are calculated in net decoupling approximation (15); dashed curves correspond to the coupling problem (13)-(14); solid curves are calculated in decoupling approximation (15) but with substitution $\alpha \to \alpha_P = \alpha + 3\tilde{\alpha}_{11}\langle P\rangle^2$ in the electrostriction term. STO parameters are listed in the **Table 1** and T=300 K is chosen.

It can be seen from **Fig. 2**, that the difference between the "coupled" (dashed) and the "decoupled" (dotted) curves is significant for thin films ($h \ll R_d$), since the unphysical divergence happens only in the decoupled case due to the electrostriction contribution, while concentration-strain and flexoelectric contributions calculated in the decoupling approximation still have satisfactory accuracy. All the curves ("coupled" and "decoupled") saturate and tend to the same value in the limit $h \gg R_d$.

At low potentials ($U \leq 0.1$ V) the response changes its sign approximately at $h \sim R_d$, since the flexoelectric and electrostriction contributions have different signs and different thickness dependence (as studied in detail in Section 3). At low potentials ($U \leq 0.1$ V), the decoupling approximation is applicable even for $h > 0.5R_d$. For high potentials, the decoupling approximation (dotted curves) works adequately at film thickness $h \gg R_d$. The accuracy of the decoupling approximation is almost independent of the value of the gradient coefficient g and carrier concentration $n_0=N_{d0}$ (compare plots **a-d** in **Fig.2**). We expect that the net linear decoupling approximation (15) should work much better for linear dielectrics.

It can be seen from the figure, that the accuracy of the decoupling approximation (15) for intermediate and high potentials can be essentially improved (up to several % from exact solution) by the substitution $\alpha \to \alpha_P = \alpha + \tilde{\alpha}_{11}\langle P\rangle^2$ in the ***electrostriction contribution*** $u_{ES} \sim \dfrac{U^2}{\alpha_P^2}\dfrac{q_{11}}{c_{11}}$ (see dotted and solid curves, which almost coincide). Note, that the renormalization accounts for the effect of paraelectric nonlinearity (nonzero $\alpha_{11}$) that is relatively strong for quantum paraelectrics like STO. Being more rigorous, one could substitute $\alpha \to \tilde{\alpha}$ and $R_d \to \tilde{R}_d$ in the flexoelectric contribution $u_{FLEXO}^{decoupled} = \dfrac{f_{11}}{c_{11}}\dfrac{U}{\alpha}\dfrac{h}{2R_d^2}$ in Eq.(15), but the substitution does not lead to any essential improvements of its accuracy, required at small thicknesses, since $\tilde{\alpha}\tilde{R}_d^2 = \frac{k_B T \tilde{\alpha}}{2N_{d0}^+ e^2}\left(\varepsilon_0 \varepsilon_b + \frac{1}{\tilde{\alpha}}\right) \approx \frac{k_B T}{2N_{d0}^+ e^2} \approx \alpha R_d^2$ for STO material parameters. The flexoelectric term in Eq.(15) coincides with the corresponding term in Eq.(13a), $u_{FLEXO}^{coupled} = \frac{f_{11}}{c_{11}}\frac{U}{\tilde{\alpha}}\frac{h^3}{2\tilde{R}_d^2(12\tilde{g}\kappa + h^2)}$, in the limit $12\tilde{g}\kappa \ll h^2$. The concentration-strain contribution $u_{CS} = n_0 \frac{eU}{k_B T}\frac{(\Xi - \beta)}{c_{11}}\frac{h}{2}$ is identical for coupled and decoupled cases at small film thicknesses.

To summarize, the difference between the electromechanical response calculated in the "coupling" and the "decoupling approximation" is very significant for thin films ($h \ll R_d$), since unphysical divergence happens only in the decoupled case due to the electrostriction contribution. Both "coupled" and "decoupled" responses saturate and tend to the same value in the limit $h \gg R_d$. The renormalization $\alpha \to \alpha_P = \alpha + \tilde{\alpha}_{11}\langle P\rangle^2$ significantly improves the linear decoupling approach for paraelectric thin films and transfers all the nonlinearity into the coefficient $\alpha_P$ that becomes $h$ and $U$ dependent. Correspondingly, this approach will be used in Section 3. Finally, for thick films with $h \gg R_d$, and especially for semi-infinite samples ($h \to \infty$), the average polarization $\langle P\rangle^2$ is zero for non-polar materials (since the electric field from the planar electrode vanishes at distances $r \gg R_d$) and $\alpha_P \equiv \alpha$, $\tilde{\alpha} \equiv \alpha$. Thus the net decoupling approximation has the appropriate accuracy for a thick film and semi-infinite samples (in agreement with Eq.(14)). Repeating similar analyses for a localized tip electrode, again leads to $\langle P\rangle^2 \to 0$ in the limit $h \to \infty$. This allows us to use the net decoupling approximation in Section 4, where the tip electrode and the limit $h \to \infty$ will be considered.

### 3. Electromechanical response of thin film in the planar capacitor geometry

For the strain measurements in the planar capacitor geometry, the top electrode is considered to be mechanically free (e.g. ultra-thin, or liquid, or soft polymer). For the cubic symmetry film on a **thick substrate,** mechanical boundary conditions are $u_i(h) = 0$, $\sigma_{i3}(0) = 0$. We also set $u_{11}(z) = u_{22}(z) = 0$, and derive the surface displacement in the form [95]:

$$u_3(0) = \int_0^h dz \left( \frac{1}{c_{11}} \left( \Xi(n(z) - n_e) + \beta(N_d^+(z) - N_{de}^+) \right) - \frac{1}{\alpha} \frac{f_{11}}{c_{11}} \frac{d^2\varphi}{dz^2} + \frac{1}{\alpha_P^2} \frac{q_{11}^{MT}}{c_{11}} \left( \frac{d\varphi}{dz} \right)^2 \right) \quad (16a)$$

To improve the accuracy of the decoupling approximation, including the effects of (paraelectric or ferroelectric) polarization nonlinearity, we substitute $\alpha \to \alpha_P = \alpha + \tilde{\alpha}_{11} \langle P \rangle^2$ and derive:

$$\alpha_P(h, U) = \alpha + \tilde{\alpha}_{11} p^2(h, U) \left( 1 - \frac{\alpha}{3\tilde{\alpha}_{11} p(h, U)^2} \right)^2 \quad (16b)$$

where the function $p(h, U) = \left( \frac{U}{2\tilde{\alpha}_{11} h} + \sqrt{\left( \frac{\alpha}{3\tilde{\alpha}_{11}} \right)^3 + \left( \frac{U}{2\tilde{\alpha}_{11} h} \right)^2} \right)^{1/3}$, the total potential $U = V_0 + V_b$, and $\tilde{\alpha}_{11} = \alpha_{11} - 2q_{11}^2 / c_{11}$.

The net (or linear) decoupling approximation is valid for low potentials $U$ and/or thick films ($h \gg R_d$). In the case where the concentration-strain and flexoelectric contributions are linear with respect to electrostatic potential $\varphi$, and thus they mimic piezoelectric effect, while electrostriction and Maxwell stress contribution are quadratic with respect to potential.

In the Debye approximation and 1D case ($V_0(r) = $ const) the solution (11) is reduced to:

$$\varphi(z) = \begin{pmatrix} e \dfrac{(N_{d0}^+ - n_0) R_d^2}{\varepsilon_0 \varepsilon} \left( 1 - \dfrac{\exp((h-z)/R_d) + \exp(z/R_d)}{1 + \exp(h/R_d)} \right) \\ + U \dfrac{\exp(-z/R_d) - \exp(-(2h-z)/R_d)}{1 - \exp(-2h/R_d)} \end{pmatrix}, \quad (17a)$$

$$N_d^+ = N_{d0}^+ \left( 1 - \frac{e\varphi}{k_B T} \right) \quad \text{and} \quad n = n_0 \left( 1 + \frac{e\varphi}{k_B T} \right). \quad (17b)$$

Within the section $n_0 \equiv n_e$ for electron-conducting interfaces, $n_0 = n_e h \Big/ \int_0^h \exp\left( \frac{e\varphi(z)}{k_B T} \right) dz$ for electron-blocking interfaces in accordance with Eq.(5b); $N_{d0}^+ \equiv N_{de}^+$ for ion-conducting interfaces and $N_{d0}^+ = N_{de}^+ h \Big/ \int_0^h \exp\left( \frac{-e\varphi(z)}{k_B T} \right) dz$ for ion-blocking interfaces in accordance with Eq.(5a). The electric potential is given by Eq.(14a).

Then direct integration of Eq.(16) leads to the analytical expression for the surface displacement:

$$u_3(z=0) = u_{CS}(0) + u_{FLEXO}(0) + u_{ES}(0). \qquad (18a)$$

Where the concentration-strain contribution depends on the interfaces (blocking/conducting) type as follows:

$$u_{CS}(0) = \begin{cases} \dfrac{\Xi n_0 - \beta N_{d0}^+}{c_{11}} W_{CS}, & \text{conducting interfaces } (\mathbf{CI}), \\[6pt] \dfrac{\Xi n_0}{c_{11}} W_{CS}, & \text{ion} - \mathbf{BI}, \text{electron} - \mathbf{CI}, \\[6pt] \dfrac{-\beta N_{d0}^+}{c_{11}} W_{CS}, & \text{ion} - \mathbf{CI}, \text{electron} - \mathbf{BI}, \\[6pt] 0, & \text{blocking interfaces } (\mathbf{BI}). \end{cases} \qquad (18b)$$

Where concentration strain contribution is

$$W_{CS} = \frac{eR_d}{k_B T}\left(\frac{e}{\varepsilon_0 \varepsilon}(N_{d0}^+ - n_0)R_d^2\left(y - 2\tanh\left(\frac{y}{2}\right)\right) + U\tanh\left(\frac{y}{2}\right)\right) \qquad (18c)$$

flexoelectric contribution is:

$$u_{FLEXO}(0) = \frac{1}{\alpha}\frac{f_{11}}{c_{11}}\frac{1}{R_d}\left(\frac{2e}{\varepsilon_0 \varepsilon}(N_{d0}^+ - n_0)R_d^2 - U\right)\tanh\left(\frac{y}{2}\right), \qquad (18d)$$

electrostriction contribution is:

$$u_{ES}(0) = \frac{1}{\alpha_P^2}\frac{q_{11}^{MT}}{c_{11}}\frac{1}{4R_d}\sinh^{-2}(y) \times$$

$$\times \left( \begin{array}{l} \dfrac{4e}{\varepsilon_0 \varepsilon}(N_{d0}^+ - n_0)R_d^2\left(U - \dfrac{e}{\varepsilon_0 \varepsilon}(N_{d0}^+ - n_0)R_d^2\right)(y\cosh(y) + \sinh(y)) + \\[6pt] \left(2\left(\dfrac{e}{\varepsilon_0 \varepsilon}\right)^2(N_{d0}^+ - n_0)^2 R_d^4 - \dfrac{2e}{\varepsilon_0 \varepsilon}(N_{d0}^+ - n_0)R_d^2 U + U^2\right)(2y + \sinh(2y)) \end{array} \right) \qquad (18e)$$

and $y = h/R_d$. It can be seen from Eqs.(18b-e) that the size effect of the contributions scales with $h/R_d$ as a relevant length scale.

The contributions to electromechanical response of STO films, namely chemical expansion/deformation potential, flexoelectric strain and electrostriction, are shown in **Figs.3** for fixed potential, ion-conducting interfaces and different concentrations of donors.

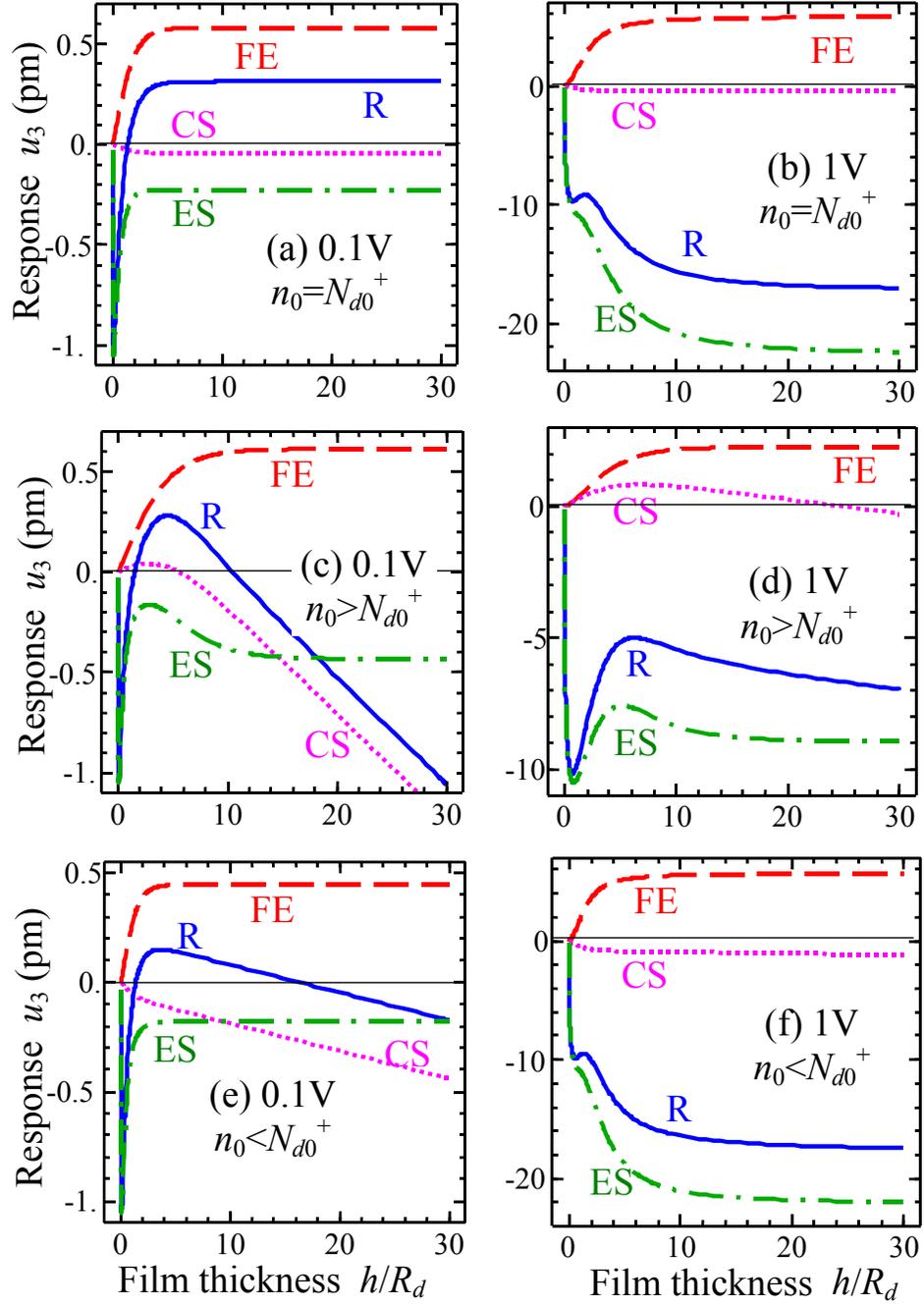

**Fig.3.** Thickness dependence of one-dimensional electromechanical response (**R**) and its concentration-strain (**CS**), electrostriction (**ES**) and flexoelectric (**FE**) contributions calculated for STO film at applied voltage $V_0$=0.1 V (a, c, e), 1 V (b, d, f), $n_0=N_{d0}=10^{25}$ m$^{-3}$ (a, b); $n_0=10^{25}$ m$^{-3}$ and $N_{d0}=10^{24}$ m$^{-3}$ (c, d); $n_0=10^{24}$ m$^{-3}$ and $N_{d0}=10^{25}$ m$^{-3}$ (e, f). STO parameters are listed in the **Table 1,** T=300 K and built-in potential $V_b$=0. The chosen space charge concentrations are $R_d \approx 4.6$ nm for the case $N_{d0}^+ = n_0$ and $R_d = 6.2$ nm for $N_{d0}^+ \neq n_0$.

The electromechanical response behaviour with increasing thickness also depends on the relationship between the donor concentration $N_{d0}^+$ and the electron concentration $n_0$ (compare **Figs. 3a,b** calculated for $N_{d0}^+ = n_0$ with **Figs. 3c-d** calculated for and $N_{d0}^+ > n_0$ and **Figs. 3e,f** calculated for $N_{d0}^+ < n_0$). Note that the condition $N_{d0}^+ \neq n_0$ may readily be achieved for thin films with planar ohmic electrodes, when the whole system "film + electrodes" remains electro-neutral due to the free charges accumulated at the planar electrodes. However, the condition $N_{d0}^+ = n_0$ holds rigorously only for a semi-infinite material, while for the films of small thickness ($h/R_d < 100$) the situation $N_{d0}^+ \neq n_0$ is possible in thin films due to the carrier injection from the planar electrodes.

It can be seen from **Figs. 3a,b** that the concentration-strain contribution to the electromechanical response is smallest under the condition $N_{d0}^+ = n_0$ for chosen material parameters. It monotonically increases with the film thickness for the case $N_{d0}^+ = n_0$.

Under the condition $N_{d0}^+ \neq n_0$ the concentration-strain contribution increases linearly with the film thickness $h$ and for small potentials $U$ may become more significant than the electrostriction and flexoelectric contributions, **Figs. 3c,e**. However, for higher $U$ the concentration-strain contribution is smaller than the electrostriction and flexoelectric contributions even at $N_{d0}^+ \neq n_0$ (see **Figs. 3d,f**). For the case $N_{d0}^+ < n_0$ the maxima exists on the thickness dependence of the concentration-strain response.

The flexoelectric contribution increases monotonically with the film thickness and saturates at $h/R_d \gg 1$ independent of the ratio between $N_{d0}^+$ and $n_0$. Both concentration-strain and flexoelectric contributions monotonically tend to zero at $h \to 0$. The electrostriction contribution has a maximum at small film thicknesses, but remains finite and even tends to zero as at $h \to 0$ allowing for the renormalization $\alpha \to \alpha_P = \alpha + \tilde{\alpha}_{11}\langle P \rangle^2$ (as shown in **Figs.3**). Note, that in net decoupling approximation, electrostriction contribution diverges at $h \to 0$ and strongly dominates at small thicknesses, since it is inversely proportional to film thickness $h$ in accordance with Eq.(18e).

The potential dependence of the STO film electromechanical response is shown in **Fig. 4** for different film thickness, potential window and concentrations of donors. It is seen that for case $N_{d0}^{+} = n_0$ the response is zero for zero potential (all the curves start from the coordinate origin), while for $N_{d0}^{+} \neq n_0$ there is a response from the bulk charge, independent of potential (compare **Fig. 4a,c** with **e,f**).

Note that all curves in **Figs.4** demonstrate a crossover from the linear ($u_3 \sim U$) at very small total potentials ($|U| \ll hE_c$), to the quasi-parabolic-like $u_3 \sim U^2$ dependences at low potentials $|U| \leq hE_c$ to the sub-linear $u_3 \sim U^{2/3}$ with the potential increase $|U| \gg hE_c$ (see log-log plots **b** and **d**). At a fixed potential window, the electromechanical response voltage dependence changes from quasi-linear and quadratic to the one with thickness decrease (**Figs. 4a,c,e,f**). This behavior can be understood as follows. At film thickness higher than the characteristic thickness $h \gg |U|/E_c$ (or at very low $|U| \ll hE_c$) the nonlinearity in equation $\left(\alpha + \tilde{\alpha}_{11}\langle P_3 \rangle^2\right)\langle P_3 \rangle = U/h$ can be disregarded and the average polarization is $\langle P_3 \rangle = U/(\alpha h)$ in accordance with Eqs.(14). Hence, the response is determined by the flexoeffect at very small potentials $U$ and is a linear function of $U$: $u_3 \sim u_{FLEXO} \sim h\langle P_3 \rangle \sim U$. At higher voltages $|U| \leq hE_c$ (but still enough low to account for nonlinearity in equation $\left(\alpha + \tilde{\alpha}_{11}\langle P_3 \rangle^2\right)\langle P_3 \rangle = U/h$) the response is determined by electrostriction and thus becomes a quadratic function of potential: $u_3 \sim u_{ES} \sim h\langle P_3^2 \rangle \sim U^2$. Finally, at high total potential $|U| \gg hE_c$ (or very small film thickness $h \ll |U|/E_c$) the nonlinearity cannot be disregarded and the dependence of polarization on potential changes to a nonlinear relationship: $\langle P_3 \rangle = (U/(\tilde{\alpha}_{11}h))^{1/3}$. As a result, the electromechanical response is determined by the biggest electrostriction contribution $u_3 \sim u_{ES} \sim h\langle P_3^2 \rangle \sim h^{1/3}(U/(\tilde{\alpha}_{11}))^{2/3}$, since flexoelectric and concentration-strain contributions increase with the potential at a much slower rate: $u_{FLEXO,CS} \sim h\langle P_3 \rangle \sim (U/(\tilde{\alpha}_{11}h))^{1/3}$ (weak sub-linear low).

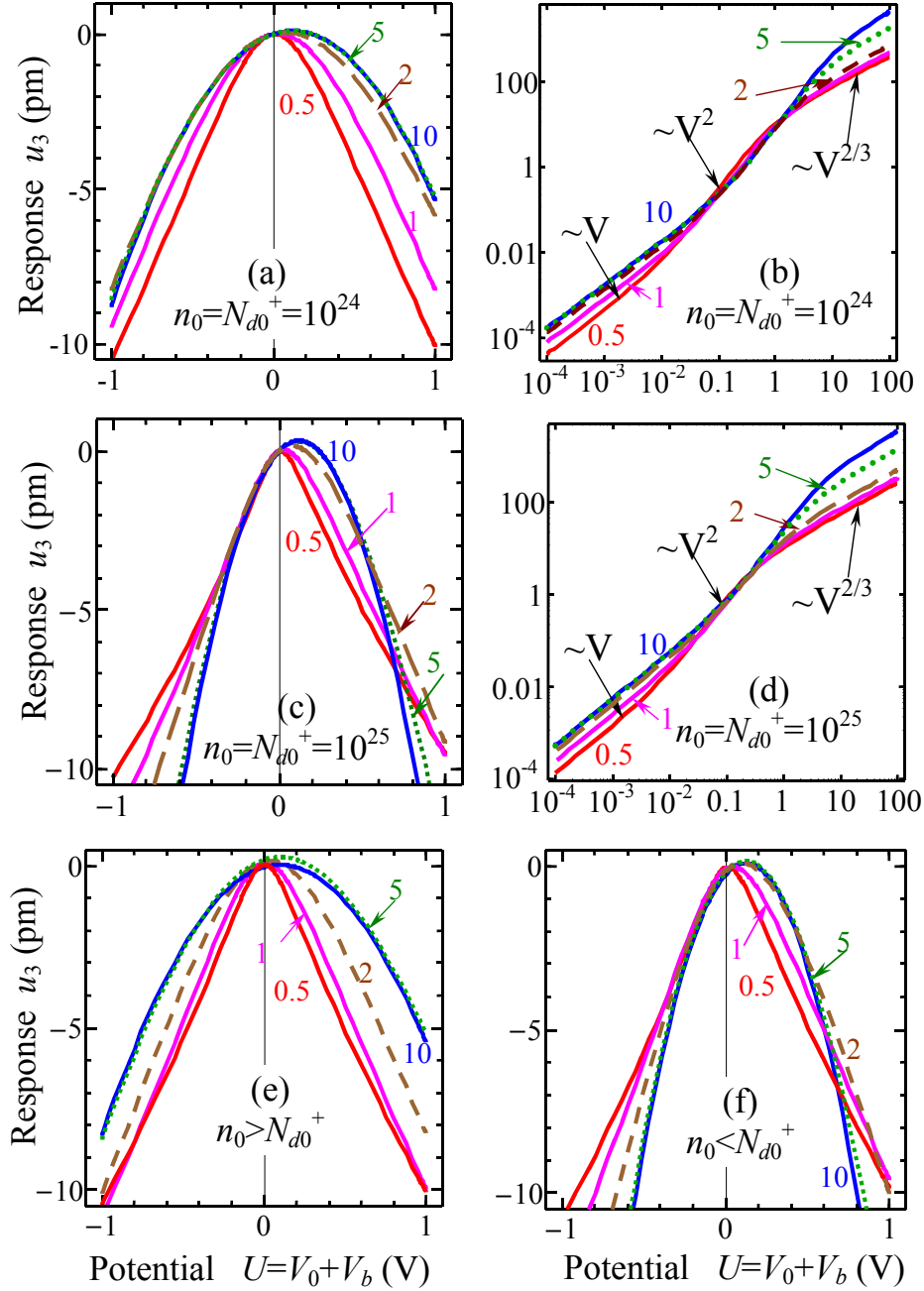

**Fig.4.** Potential dependence of the one-dimensional electromechanical response of STO film of thickness $h/R_d = 1, 2, 5, 10, 20$ (figures near the curves), $n_0=N_{d0}=10^{24}$ m$^{-3}$ (a – linear scale, b –

log-log scale); $n_0=N_{d0}=10^{25}$ m$^{-3}$ (c – linear scale, d – log-log scale); $n_0=10^{25}$ m$^{-3}$ and $N_{d0}=10^{24}$ m$^{-3}$ (e), $n_0=10^{24}$ m$^{-3}$ and $N_{d0}=10^{25}$ m$^{-3}$ (f). STO parameters are listed in the **Table 1** and T=300 K.

Contour plot of the electromechanical response in coordinates total potential – film thickness is presented in **Fig. 5** for the case $N_{d0}^+ = n_0$.

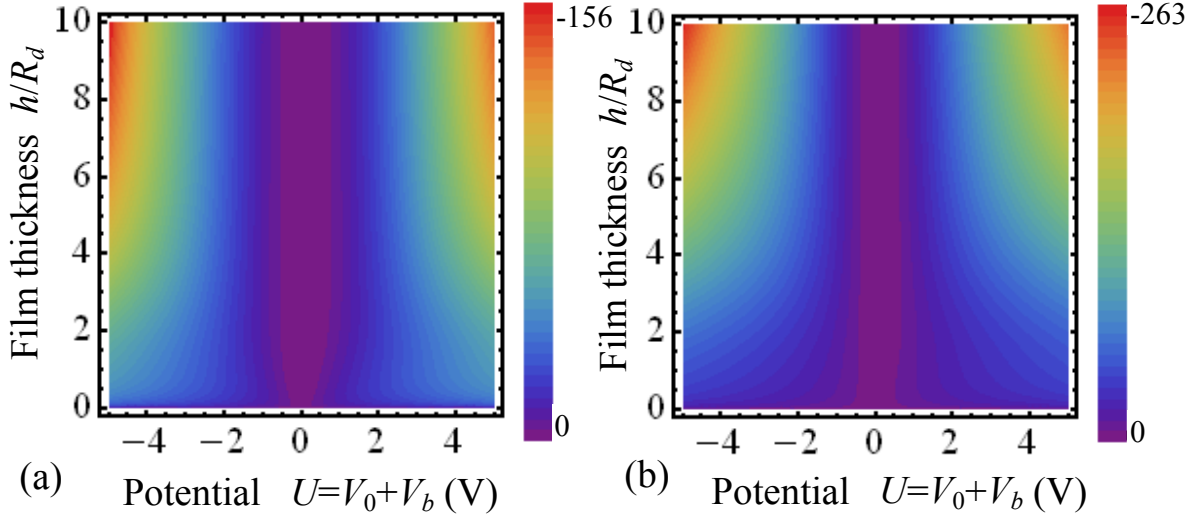

**Fig.5.** Contour plot of the electromechanical response in coordinates total potential – film thickness. Color scales indicate the response values in picometers (pm=10$^{-12}$m). STO parameters are listed in the **Table 1,** $n_0=N_{d0}=10^{24}$ m$^{-3}$ (a) and $n_0=N_{d0}=10^{25}$ m$^{-3}$ (b); also we set T=300 K.

It can be seen from **Fig. 5** that the shape of the contours is asymmetric with respect to the sign of total potential $U$ at small thicknesses. The asymmetry originates from flexoelectric and concentration-strain contributions $u_{CS,FLEXO}$, which are odd functions of the total potential $U$ at $N_{d0}^+ = n_0$, while the electrostriction contribution $u_{ES}(0)$ is the even function of $U$ at $N_{d0}^+ = n_0$. The asymmetry vanishes with increasing film thickness. In the decoupling approximation, the electrostriction contribution has a quadratic potential dependence and dominates with potential increase $u_{ES}(0) \sim AU^2 + BU + C_1$ (parabolic form of some curves), while other contributions have linear dependence, $u_{CS,FLEXO}(0) \sim DU + C_2$, and responsible for the asymmetry of the curves, since the coefficients $C_{1,2} \sim (N_{d0}^+ - n_0)$ and $B \sim (N_{d0}^+ - n_0)$ are zero in the case $N_{d0}^+ = n_0$.

To summarize, the electromechanical response of SrTiO$_3$ films exhibits a size effect controlled by the ratio of the film thickness $h$ to the carriers screening radius $R_d$, and all the contributions (electrochemical, flexoelectric, electrostriction and Maxwell stain) of the response depend on the ratio $h/R_d$ and saturate at $h/R_d \gg 1$. The potential dependence of the response demonstrates a crossover from the linear to the quadratic law and then to the sub-linear (power law 2/3) with a decrease in the film thickness at a fixed potential window (or with increasing applied voltage and fixed film thickness). The crossover originates from the cubic nonlinearity in electric polarization field dependence, typical for paraelectrics.

## 4. Local electromechanical response for PFM geometry

For the radius-dependent potential $V_0(r)$ induced by the SPM probe, the local displacement of the STO surface given by Eq.(12a) is as follows:

$$u_3(z=0,r) = u_{CS}(r) + u_{FLEXO}(r) + u_{ES}(r) + u_3^{PL}, \tag{19a}$$

Given that **decoupling approximation** has good accuracy for a thick film and semi-infinite samples (see section 2.4), below we will consider an elastically isotropic thick film with thickness $h/R_d \gg 1$ and $h/R_0 \gg 1$. In order to consider another limiting case $h/R_0 \ll 1$ one can easily use the results of the previous section and obtain at least semi-quantitative results. Furthermore, we assume $N_{d0}^+ \approx n_0$ as expected for thick films, but locally $N_d^+(\mathbf{r}) \neq n(\mathbf{r})$, especially in the depletion/accumulation regions (see Eqs.(4)).

However, the decoupling approximation assumes that the average polarization $\langle P_i \rangle$ induced by the applied voltage is sufficiently small in the actual region of electromechanical response, so that its nonlinear voltage dependence in the equation $(\alpha + \tilde{\alpha}_{11} \langle P_i^2 \rangle) \langle P_i \rangle = \langle E_i \rangle$ can be disregarded across the entire signal generation volume. Since the "bare" tip electric field is Coulombic, the actual region of electromechanical response is proportional to the semi-sphere with radius $R_{PFM} \sim 10 \cdot \max\{R_0, R_d\}$ for the considered thick film with thickness $h/R_0 \gg 1$ and $h/R_d \gg 1$. Using the results of Section 2.4, we do not consider the average polarization nonlinearity in the response region at the voltages $|V_0| \leq R_{PFM} E_c$, where the field $E_c$ is given by Eq.(14b). For STO parameters at room temperature it gives $|V_0| < (4-20)\,\text{V}$ at $R_0 = (10-50)\,\text{nm}$.

Using the decoupling approximation and the Debye approximation (11) for electrostatic potential in **Appendix B** we obtained the closed-form expressions for concentration-strain, flexoelectric and electrostriction (renormalized by Maxwell stress) contributions in semi-infinite sample limit $h \to \infty$ as:

$$u_{CS}(r) = \frac{2(1+\nu)(1-2\nu)}{Y}(\Xi - \beta)\frac{N_{d0}^{+}e}{k_B T}\int_0^\infty dk J_0(kr)\frac{k \cdot \tilde{V}_0(k)}{K(k)+k}, \quad (19b)$$

$$u_{FLEXO}(r) = -\frac{(1+\nu)}{R_d^2 Y}\int_0^\infty dk J_0(kr) k \tilde{V}_0(k)\left(\begin{array}{c} \dfrac{f_{11}+f_{12}}{\alpha}\dfrac{(1-2\nu)}{(k+K(k))} + \\ \dfrac{f_{11}-f_{12}}{\alpha}\dfrac{k}{(k+K(k))^4 R_d^2} \end{array}\right) \quad (19c)$$

$$u_{ES}(r) \approx \frac{(1+\nu)}{Y}\int_0^\infty dk J_0(kr) k \left(\begin{array}{c} \dfrac{q_{11}^{MT}+q_{12}^{MT}}{\alpha^2}(1-2\nu)\left(\dfrac{2K(k)\tilde{V}_0(k)V_b}{1+(k+K(k))R_d} + \dfrac{\tilde{V}_0^2(k/2)}{2R_0^2(k+2K(k/2))R_d^2}\right) \\ + \dfrac{q_{11}^{MT}-q_{12}^{MT}}{\alpha^2}\left(\dfrac{2kR_d(k+K(k))\tilde{V}_0(k)V_b}{(1+(k+K(k))R_d)^2} + \dfrac{\tilde{V}_0^2(k/2)4k}{2R_0^2(k+2K(k/2))^4 R_d^4}\right) \end{array}\right)$$

(19d)

Here $Y$ denotes the Young module, $\nu$ the Poisson ratio, $K(k) = \sqrt{k^2 + R_d^{-2}}$ and polar radius $r = \sqrt{x_1^2 + x_2^2}$, $J_0(x)$ the Bessel function of the zero order and $\tilde{V}_0(k)$ the x,y-Fourier transform of the applied potential $V_0(r)$. Equation (19d) is valid for well-localized probe potentials $V_0(r)$, when the integration can be performed by the Laplace method. Note, that $n_0 \equiv n_e = N_{d0}^{+} \equiv N_{de}^{+}$ for a semi-infinite sample due to the electric field and strain vanishing at depth $z \gg R_d$. The PFM tip has finite size and is regarded as electron-conductive, but can be both ion-blocking or ion-conducting; the surface outside the contact area is ionically conductive (e.g. can support ion-exchange reaction with gas or liquid phase).

The expression for the radius-independent displacement $u_3^{PL} \sim V_b$, which originates from the constant built-in potential, can be derived similarly to Eqs.(18) in the limit $h \to \infty$, when $N_{d0}^{+} = n_0$. Namely:

$$u_3^{PL} = V_b \frac{(1+\nu)(1-2\nu)}{Y}\left(\frac{e(\Xi-\beta)n_0}{k_B T}R_d - \frac{f_{11}+f_{12}}{\alpha R_d} + \frac{q_{11}^{MT}+q_{12}^{MT}}{\alpha^2}\frac{V_b}{2R_d}\right). \quad (19e)$$

To define the boundary condition representative of the ESM experiment, we assume that the tip potential $V_0(r)$ is applied inside the circle of radius $R_0$ and rapidly decays outside. For numerical estimations, we used the **Gaussian model** for the potential and its x,y-Fourier image of the surface, namely:

$$V_0(r) = V_0 \exp\left(-\frac{r^2}{2R_0^2}\right), \qquad \tilde{V}_0(k) = V_0 R_0^2 \exp\left(-\frac{(kR_0)^2}{2}\right). \tag{20}$$

Here $R_0$ is either the tip effective size or the tip-surface contact radius and $V_0$ is the voltage applied to the tip.

The model potential (20) allows integration in general expressions (19b-d) leading to the Pade-approximations (see also exact analytical expressions in **Appendix C**) for the maximal surface displacement at $r=0$:

$$u_{CS}(0) \approx \frac{2(1+\nu)(1-2\nu)}{Y}(\Xi-\beta)N_{d0}^+ \frac{V_0 e}{k_B T} \frac{R_d y}{\sqrt{(8/\pi)+\sqrt{2\pi}y+y^2}} \tag{21a}$$

$$u_{FLEXO}(0) \approx -\frac{(1+\nu)}{Y}\frac{V_0}{\alpha R_d}\left(\frac{(1-2\nu)(f_{11}+f_{12})y}{\sqrt{(8/\pi)+\sqrt{2\pi}y+y^2}} + \frac{(f_{11}-f_{12})y^3}{13+13y+5y^2+\sqrt{(2/\pi)}y^3}\right) \tag{21b}$$

$$u_{ES}(0) \approx \frac{(1+\nu)}{YR_d}\left(\begin{array}{c}\dfrac{q_{11}^{MT}+q_{12}^{MT}}{\alpha^2}(1-2\nu)\left(V_0 V_b\left(1-\sqrt{\dfrac{\pi}{8}}\dfrac{y}{(1+y)^2}\right)+\dfrac{V_0^2 y}{2\sqrt{(4/\pi)+\sqrt{\pi}y+y^2}}\right)\\ +\dfrac{q_{11}^{MT}-q_{12}^{MT}}{\alpha^2}\left(\dfrac{V_0 V_b}{\sqrt{1+\sqrt{2\pi}y+(8/\pi)y^2}}+\dfrac{V_0^2 y^2}{13+19y+10y^2+(4/\sqrt{\pi})y^3}\right)\end{array}\right) \tag{21c}$$

Here the introduced the dimensionless tip size $y = R_0/R_d$. Accuracy of the expressions (21) was analyzed numerically and it appeared not less than several percent. It can be seen from these expressions that the concentration-strain, flexoelectric, flexoelectric contributions to the PFM response scale are given by $y = R_0/R_d$.

Using expressions (19)-(20) we calculated the dependence of the PFM response $u_3(0,0) - u_3^{PL}$ on the dimensionless tip size $y = R_0/R_d$, as shown in **Fig.6** for different built-in potentials $V_b=0$, $V_b>0$, $V_b<0$. From **Figs.6**, one can observe that the electrostriction contribution to PFM response dominates for applied voltages $V_0>1$ V, when flexoelectric and concentration-strain contributions are smaller in one and two orders of magnitude for chosen material parameters. Additional calculations show that flexoelectric and concentration-strain contributions overcome the electrostriction contribution for small built-in potentials $|V_b| \leq 0.1$ V, applied voltages $|V_0| \leq 0.1$ V and room temperature (compare with **Figs. 3** plotted for $V_0 = 0.1$ and 1 V).

The electrostriction contribution (and consequently total PFM response) increases with the tip size for positive applied voltage $V_0 > 0$ and positive (or zero) built-in potentials $V_b$, then reaches a saturation point and becomes almost independent of the tip size at $R_0 > 10 R_d$. For $V_b < 0$ and $V_0 > 0$ it decreases with increasing tip size, then saturates and becomes almost independent of the tip size at $R_0 > 10 R_d$. The reverse situation holds for $V_0 < 0$. Analytical expressions Eqs. (21) proved that the PFM response contributions monotonically increase with the tip size at $V_0 V_b \geq 0$. The concentration-strain and flexoelectric contributions always monotonically increase with the tip size, then saturate and become almost independent of the tip size at $R_0 > 10 R_d$.

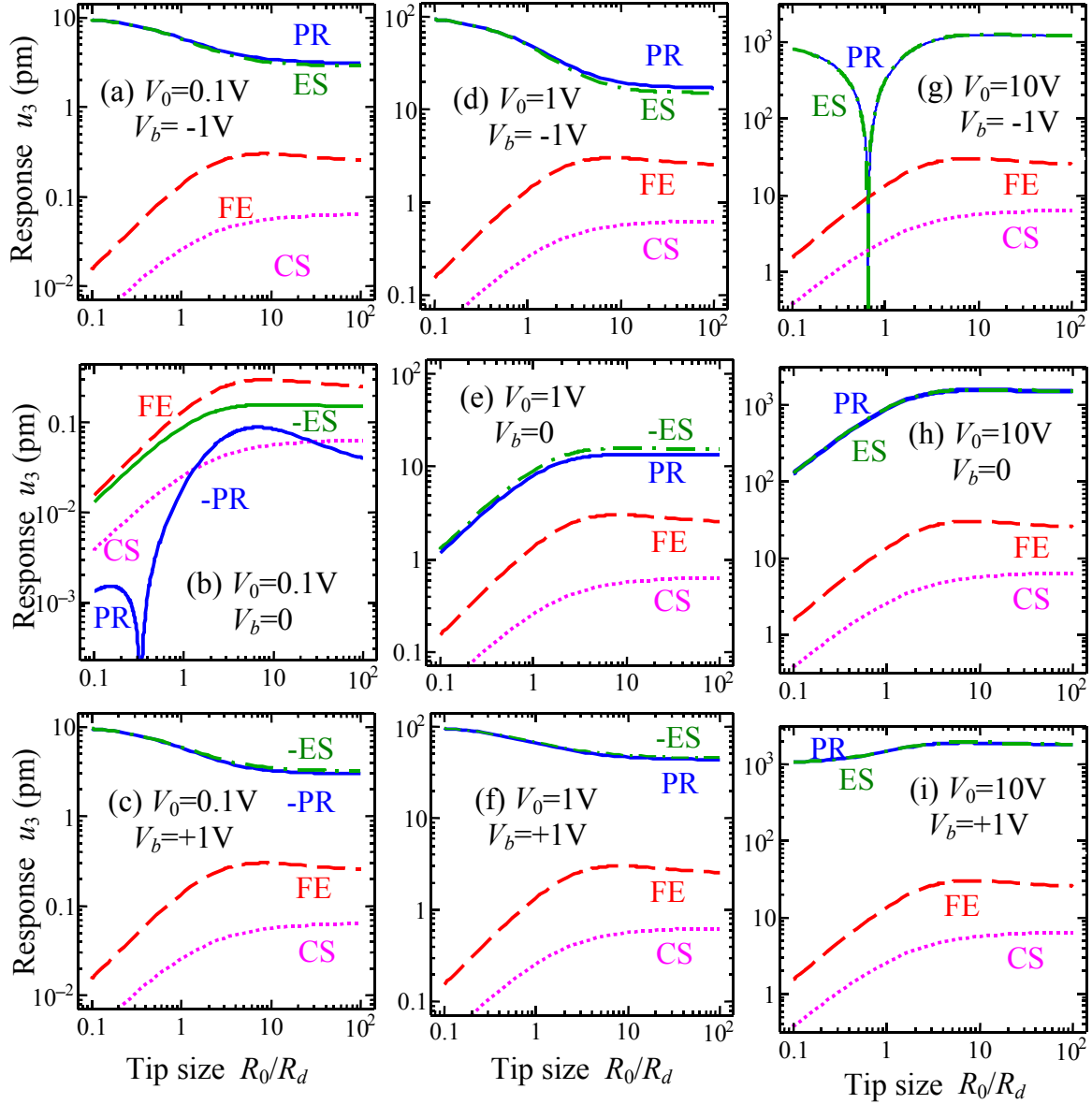

**Fig.6.** Absolute value of PFM response (**PR**) of STO vs. effective tip size $y = R_0/R_d$. The concentration-strain (**CS**), electrostriction (**ES**) and flexoelectric (**FE**) contributions are shown. Applied voltage $V_0$=0.5 V (a, b, c), 1 V (d, e, f), 10 V (g, h, i), built-in potential $V_b$=-1 (a, d, g), $V_b$=0 (b, e, h), $V_b$=1 (c, f, i). STO parameters are listed in the **Table 1,** $n_0=N_{d0}=10^{25}$ m$^{-3}$ and we set T=300 K. For chosen space charge concentrations $R_d \approx 4.6$ nm.

The voltage dependence of the STO PFM response $u_3(0,0) - u_3^{PL}$ is shown in **Fig. 7** for different tip radii and built-in potentials $V_b$=0, $V_b$<0, $V_b$>0. It can be seen from **Fig. 7b** in log-log scale, that the electrostriction contribution has quadratic voltage dependence and becomes dominant for large voltages. It can also be observed from the comparison of **Figs. 7a, c** and **d,** calculated for $V_b<0$, $V_b=0$, $V_b>0$ correspondingly, that the built-in potential leads to the horizontal shift of the parabolic curves as anticipated.

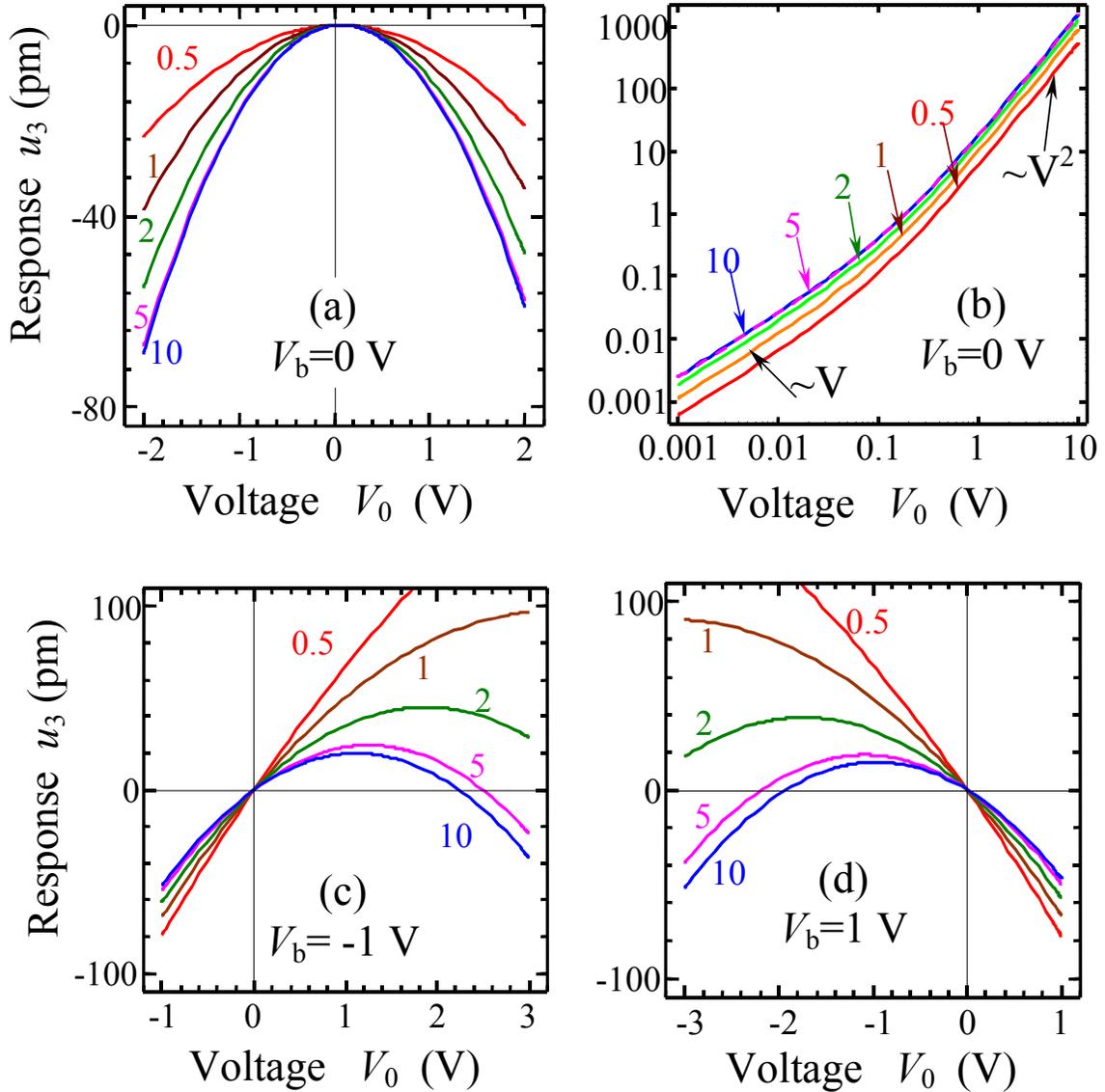

**Fig.7.** Voltage dependence of the PFM response of STO surface calculated for the tip size $R_0/R_d = 0.5, 1, 2, 5, 10$ (figures near the curves) and built-in potential $V_b=0$ (a - linear scale, b – log-log scale), $V_b<0$ (c), $V_b>0$ (c). STO parameters are listed in the **Table 1** and we set T=300 K and $n_0=N_{d0}=10^{25}$ m$^{-3}$.

Contour plot of the PFM response $u_3(0,0) - u_3^{PL}$ in coordinates voltage – tip radius is presented in **Fig. 8** for different built-in potentials $V_b<0$, $V_b=0$, $V_b>0$. Note that the shape of the contours is asymmetric with respect to the sign of the tip voltage $V_0$ at small tip radii and $V_b \neq 0$. The asymmetry vanishes with the tip radius increase.

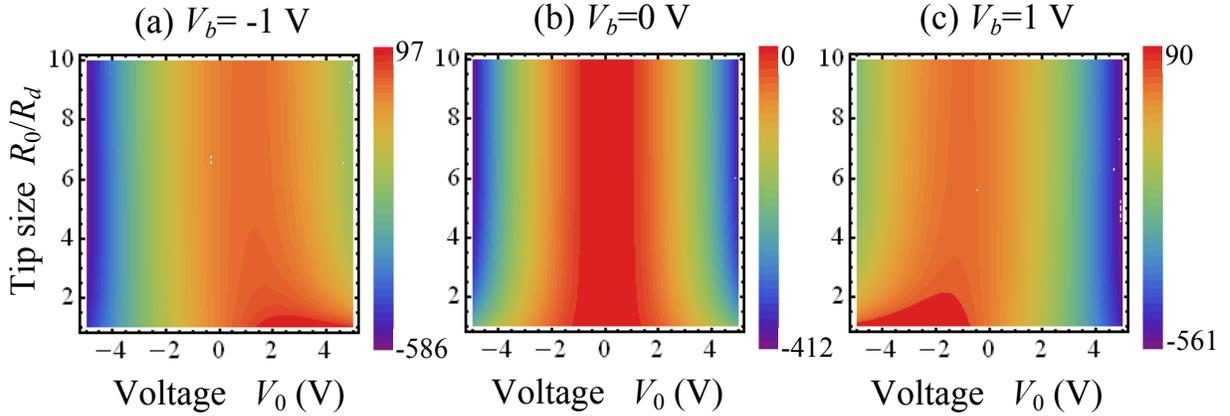

**Fig.8.** Contour plot of the STO PFM response in coordinates applied voltage – tip radius. Color scales indicate the response values in pm. Built-in potential $V_b<0$ (a), $V_b=0$ (b), $V_b>0$ (c), $n_0=N_{d0}=10^{25}$ m$^{-3}$ (b); STO parameters are listed in the **Table 1** and T=300 K.

*To summarize*, concentration-strain, flexoelectric and electrostriction contributions to the electromechanical response of the paraelectric semiconducting material depend on the tip (or contact) radius for $R_0/R_d \leq (10 - 10^2)$ (see Eqs.(21) and **Fig.6**). In comparison, piezoelectric response, that is dominant for ferroelectrics-dielectrics and ferroelectric-semiconductors and zero for considered non-polar materials, is almost independent of the tip size or contact radius $R_0$ [96, 97, 98, 99, 100, 101, 102]. This behavior stems from the difference in physical mechanisms of the response formation: piezoelectric response is linear in polarization and exists only in materials without an inversion center, the flexoelectric response is linear in the polarization gradient, electrostriction is quadratic in polarization and exist in all symmetries, and the concentration-strain contribution is linear in the changes of ionic and electronic concentrations.

## 5. Temperature dependences of quantum paraelectric PFM response

The data shown in **Figures 3-8** corresponds to room temperature. Since the dielectric permittivity $\varepsilon(T)$ of quantum paraelectric strongly increases with decreasing temperature in accordance with Barrett Law for $\alpha(T)$, we expect that all contributions $u_{CS}$, $u_{FLEXO}$ and $u_{ES}$ should be strongly temperature dependent via the temperature dependence of $\varepsilon(T)$, $\alpha(T)$ and $R_d(T)$. The electrostriction contribution may dominate at even lower voltages and low temperatures. Below, we analyze the temperature dependences of the STO surface local displacement caused by the well-localized potential $V_0(r)$ of the SPM probe. As in the previous section, we use the ***net decoupling approximation*** for elastically isotropic films with thickness $h/R_d \gg 1$ and $h/R_0 \gg 1$. Also we set $N_{d0}^+ \approx n_0$ as expected for thick films.

In the decoupling approximation, the PFM response given by Eqs.(21) is temperature dependent via the temperature dependence of $\alpha(T) = \alpha_T \left( \coth(T_q/(2T)) T_q/2 - T_0 \right)$ in accordance with Barrett Law (see **Table 1**), dielectric permittivity $\varepsilon(T) = \varepsilon_b + (\alpha(T)\varepsilon_0)^{-1}$, Debye screening radius $R_d(T) \sim \sqrt{\varepsilon(T)T/n_0(T)}$ and carrier concentration amplitude obeys the conventional activation law $n_0(T) = N_{d0}^+(T) = N_\infty \exp(-E_a/k_B T)$ [89].

The temperature dependences of the STO PFM response are shown in **Fig.9**. The decrease in the absolute value of the PFM response with the decrease in temperature could be attributed to the sharp increase of the Debye screening radius (see **Fig. 10c, d**) due to a electron concentration decrease with the fall in temperature (**Fig.10a**), while the decrease of PFM response with a temperature increase is related to the temperature dependence of coefficient $\alpha$ (see **Fig. 10b**). The response is non-monotonic due to different temperature dependences of $n_0$, $R_d$ and $\alpha$. As a result of such non-monotonic behaviour the response reveals a maximum at intermediate temperatures. The maxima temperature monotonically decreases with an increasing tip radius; its height increases and halfwidth decreases with an increasing tip radius (compare different curves calculated for different $R_0 = 3$, 10, 30, 100 nm in **Fig. 9a-c**). The absolute value of the maximal response increases with the tip voltage (compare **Fig. 9a-c**). The contour map shown in **Fig. 9d** demonstrates the possibility of observing the highest PFM response in the temperature range 300-500 K at the tip size $R_0 = 10$ nm. The temperature range of maximal response strongly depends on the tip size.

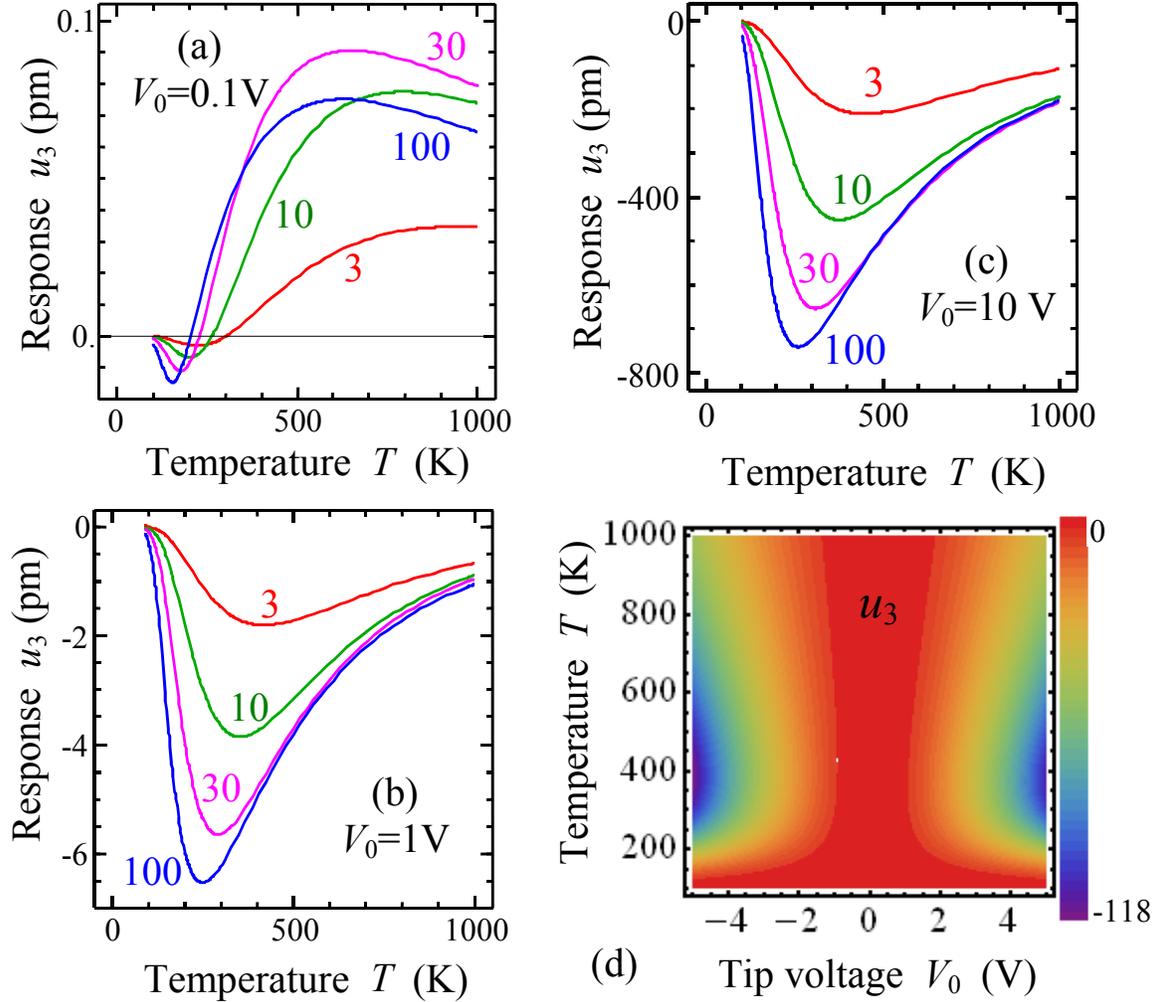

**Fig. 9.** Temperature dependence of the STO PFM response calculated at applied voltage $V_0$=0.1, 1, 10 V (a, b, c). Different curves in plots (a-c) correspond to different tip radii $R_0$ =3, 10, 30, 100 nm (figures near the curves). (d) Contour map of the STO PFM response in coordinates voltage-temperature at $R_0 = 10$ nm. Color scale indicates the response values in the units of pm. STO parameters are listed in the **Table 1**; the built-in potential is $V_b$=0 and the temperature dependence of concentrations are taken as $n_0=N_{d0}=N_\infty \exp(-E_a/(k_B T))$ with $N_\infty$=$10^{26}$ m$^{-3}$ and $E_a$=0.1 eV.

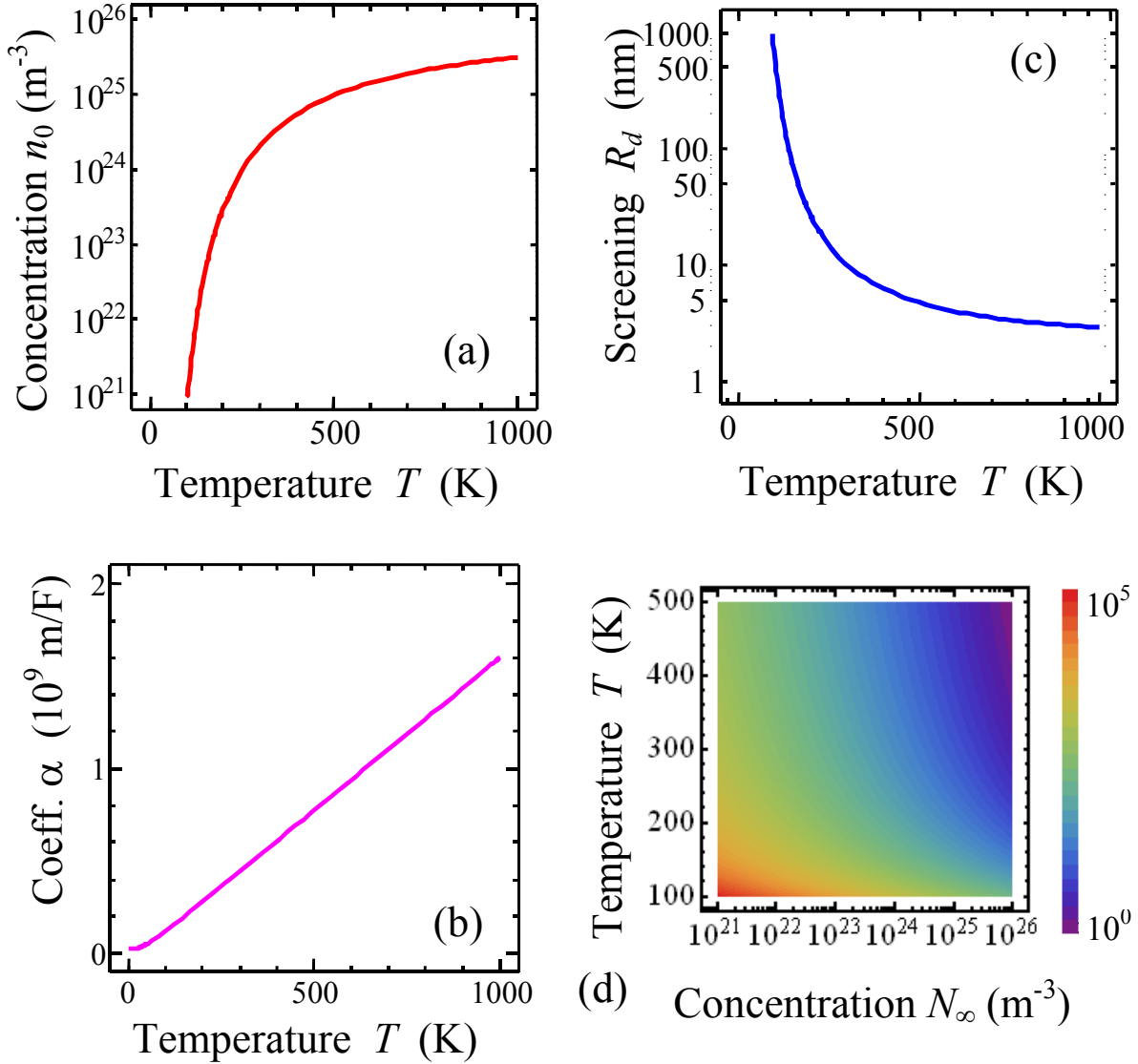

**Fig. 10.** Concentration (a), expansion coefficient (b) and screening radius (c) temperature dependences. (d) Contour map of the screening radius in coordinates concentration-temperature. Color scale indicates the radius values in nm. STO parameters are listed in the **Table 1**; the built-in potential $V_b=0$ and the temperature dependence of concentrations are taken as $n_0=N_{d0}=N_\infty \exp(-E_a/(k_B T))$ with values $N_\infty=10^{26}$ m$^{-3}$ and $E_a=0.1$ eV, following Ref.[84].

***To summarize the results of Section 5,*** the PFM response of quantum paraelectrics has pronounced temperature dependence with maxima. The maxima position, height and halfwidth, strongly depends on the tip effective size (or contact radius) at the same other parameters (e.g. at the same applied voltage and carrier concentration). Since the material parameters of quantum paraelectrics are relatively well known, the obtained analytical results can help to us to make the choice regarding the optimal experimental conditions (tip size, voltage window, working temperature, etc) to observe the maximal PFM response.

## 5. Summary

The electromechanical response of a paraelectric film to external voltage in the planar capacitor and SPM geometry is analyzed in detail including the effects of electrostriction, electrochemical coupling, and flexoelectricity. The response demonstrates a strong size effect, controlled by the ratio of the film thickness $h$ to the carriers screening radius $R_d$. The voltage dependence has a pronounced crossover from the linear to the quadratic law and then to the sub-linear. The crossover originates from the cubic nonlinearity in electric polarization field dependence, typical for quantum paraelectrics.

For the PFM configuration, the electrochemical, flexoelectric, electrostriction and Maxwell strain contributions all have different dependencies on film thickness and tip size (contact radius). The difference in the responses size dependence originates from the principal difference in the physical mechanisms of the response formation: flexoelectric response is linear in the polarization gradient, electrostriction is quadratic in polarization and exists in all symmetries, and electrochemical concentration-strain contribution is linear in the ions and electrons concentration variations. The difference between the electromechanical response calculated for the fully coupled case and in the decoupling approximation is significant for thin films ($h << R_d$), since unphysical divergence happens only in the decoupled case due to the electrostriction contribution. Both "coupled" and "decoupled" responses saturate and tend to the same value in the limit $h >> R_d$.

Finally, we predict that the PFM response of quantum paraelectrics has a pronounced non-monotonic temperature dependence. The maxima position, height and halfwidth, strongly depend on the tip effective size (or contact radius). Since the material parameters of quantum paraelectrics are relatively well known, the obtained analytical results can help one to make the choice regarding the optimal experimental conditions (tip size, voltage window, working temperature, etc) to observe the maximal PFM response.

**Acknowledgements**

One of the authors (SVK) was supported by the US Department of Energy Office of Basic Energy Sciences. ANM, EEA and GSS acknowledges financial of NAS Ukraine. Research was sponsored by in part by Ukraine State Committee on Science, Innovation and Information (UU30/004) and National Science Foundation (DMR-0908718). ANM, EEA and GSS acknowledge user agreement with CNMS N UR-08-869.

**Appendix A.**

**A1. Derivation of the electrostatic free energy**

The pure electrostatic part of (1b) can be derived from the conventional electrostatic energy

$$F_{EL} = \int_V d^3r \frac{\mathbf{DE}}{2} + \int_S d^2r \varphi \sigma_f = \left| D_n = \sigma_f \right| = \int_V d^3r \frac{\mathbf{DE}}{2} + \int_S d^2r \varphi (\mathbf{Dn}) =$$

$$= \int_V d^3r \left( \frac{\mathbf{DE}}{2} + \mathrm{div}(\varphi \mathbf{D}) \right) = \int_V d^3r \left( \frac{\mathbf{DE}}{2} + \varphi \mathrm{div}\mathbf{D} + \mathbf{D}\mathrm{grad}\varphi \right) = \left| \begin{array}{l} \mathrm{div}\mathbf{D} = \rho_f \\ \mathrm{grad}\varphi = -\mathbf{E} \end{array} \right| = \int_V d^3r \left( -\frac{\mathbf{DE}}{2} + \varphi \rho_f \right) =$$

$$= \left| \begin{array}{l} \mathbf{D} = \mathbf{P} + \varepsilon_0 \varepsilon_b \mathbf{E} \\ \mathbf{E} = \varepsilon_0^{-1} \hat{\chi}^{-1} \mathbf{P} = \hat{\alpha} \mathbf{P} \end{array} \right| = \int_V d^3r \left( -\frac{\mathbf{PE}}{2} - \frac{\varepsilon_0 \varepsilon_b \mathbf{E}^2}{2} + \varphi \rho_f \right) = \int_V d^3r \left( \frac{\alpha_{ij}}{2} P_i P_j - \mathbf{PE} - \frac{\varepsilon_0 \varepsilon_b \mathbf{E}^2}{2} + \varphi \rho_f \right)$$

Here $\sigma_f$ is the free charge accommodated at the surface. Note, that we used Gauss theorem in the derivation.

**A2. Electrostatic potential in Debye approximation**

Derivation of the solution (11)

$$\begin{cases} \Delta \varphi - \frac{\varphi}{R_D^2} = -\frac{e}{\varepsilon_0 \varepsilon} \left( N_{d0}^+ - n_0 \right), \\ \varphi(x_1, x_2, 0) = V_0(r) + V_b, \quad \varphi(x_1, x_2, h) = 0 \end{cases} \quad (A.1)$$

$$\varphi = \begin{pmatrix} e \frac{\left( N_{d0}^+ - n_0 \right) R_d^2}{\varepsilon_0 \varepsilon} + A \exp(-z/R_d) + B \exp(z/R_d) \\ + A_k \exp(-K(k)z) + B_k \exp(-K(k)z) \end{pmatrix} \quad (A.2)$$

Where $K(k) = \sqrt{k^2 + R_d^{-2}}$. Boundary conditions (5) give the equations for the constants

$$e \frac{\left( N_{d0}^+ - n_0 \right) R_d^2}{\varepsilon_0 \varepsilon} + A + B = V_b, \quad e \frac{\left( N_{d0}^+ - n_0 \right) R_d^2}{\varepsilon_0 \varepsilon} + A \exp(-h/R_d) + B \exp(h/R_d) = 0, \quad (A.3b)$$

$$A_k + B_k = \tilde{V}_0(k), \quad A_k \exp(-K(k)h) + B_k \exp(-K(k)h) = 0. \tag{A.3b}$$

## A3. Accuracy of decoupling approximation

To estimate the accuracy of decoupling approximation we studied numerically model 1D case when all variables are only z-dependent in the coupled system (2), (3), (7) using Debye approximation (10c) for electrostatic potential and regard $N_{d0}^+ = n_0$.

$$(\alpha + 2u_{33}q_{33})P_3 - g_{33}\frac{\partial^2 P_3}{\partial z^2} = f_{33}\frac{\partial u_{33}}{\partial z} + E_3, \tag{A.4a}$$

$$\varepsilon_0 \varepsilon_b \frac{\partial^2 \varphi}{\partial z^2} = \frac{2n_0 e^2}{k_B T}\varphi + \frac{\partial P_3}{\partial z}. \tag{A.4b}$$

Electric field component is $E_3 = -\partial\varphi/\partial z$.

Mechanical boundary conditions for 1D systems it means that $\sigma_{33} = 0$ everywhere, so the strain $u_{33}$ calculated from Eq.(6) acquires the form

$$u_{33}(z) = -n_0 \frac{e\varphi}{k_B T}\frac{\Xi - \beta}{c_{11}} - \frac{f_{33}}{c_{11}}\frac{\partial P_3}{\partial z} - \frac{q_{33}}{c_{11}}P_3^2 - \frac{E_3}{2c_{11}}(P_3 + \varepsilon_0 \varepsilon_b E_3) \tag{A.5}$$

Within the coupled model we minimize the free energy (1) numerically and analytically by direct variational method [72] and then derived approximate analytical solution for electric field $E_3(z)$, polarization $P_3(z)$ and strain $u_{33}(z)$ distributions. The approximate analytical dependence of the displacement (9) on applied voltage has the form:

$$u_3^{coupled}(0) = \frac{Ue}{k_B T}\frac{\Xi - \beta}{c_{11}}\frac{n_0}{\tilde{\alpha}s_1 s_2(s_1^2 - s_2^2)}\left(s_2^3(\tilde{g}s_1^2 - \tilde{\alpha})\tanh\left(\frac{s_1 h}{2}\right) - s_1^3(\tilde{g}s_2^2 - \tilde{\alpha})\tanh\left(\frac{s_2 h}{2}\right)\right) +$$

$$+ \frac{f_{11}}{c_{11}}\frac{U}{\tilde{\alpha}}\frac{s_1 s_2}{(s_1^2 - s_2^2)}\left(s_2 \tanh\left(\frac{s_1 h}{2}\right) - s_1 \tanh\left(\frac{s_2 h}{2}\right)\right) +$$

$$+ \frac{q_{11}}{c_{11}}\left(\begin{array}{l}h\langle P\rangle^2 + \left(\frac{U}{\tilde{\alpha}}\right)^2 \frac{s_1 s_2(s_2^3 \coth(s_1 h) + s_1^3 \coth(s_2 h) + s_1 s_2 h(s_2^2 \sinh(s_1 h)^{-2} + s_1^2 \sinh(s_2 h)^{-2}))}{2(s_1^2 - s_2^2)^2} \\ + \left(\frac{U}{\tilde{\alpha}}\right)^2 \frac{2s_1^3 s_2^3(s_2 \coth(s_1 h) - s_1 \coth(s_2 h))}{(s_1^2 - s_2^2)^3} - \frac{1}{h}\left(\frac{U}{\tilde{\alpha}}\right)^2\end{array}\right)$$

$$\tag{A.6}$$

Here $U = V_0 + V_b$. Spatial scales

$$s_{1,2} = \sqrt{\frac{1}{2}\left(\frac{1}{R_D^2} + \frac{1}{\kappa\widetilde{g}} + \frac{\widetilde{\alpha}}{\widetilde{g}} \pm \sqrt{\left(\frac{1}{R_D^2} + \frac{1}{\kappa\widetilde{g}} + \frac{\widetilde{\alpha}}{\widetilde{g}}\right)^2 - \frac{4}{R_D^2}\frac{\widetilde{\alpha}}{\widetilde{g}}}\right)} \quad (A.7)$$

Here we introduced coefficient $\widetilde{\alpha}_{11} = \alpha_{11} - 2q_{11}^2/c_{11}$ renormalized by electrostriction, inverse susceptibility $\widetilde{\alpha} = \alpha + 3\widetilde{\alpha}_{11}\langle P\rangle^2$ renormalized by paraelectric nonlinearity, permittivity $\kappa = \varepsilon_0\varepsilon_b$, Debye screening radius $\widetilde{R}_d^2 = \frac{k_B T}{2N_{d0}^+ e^2}\left(\varepsilon_0\varepsilon_b + \frac{1}{\widetilde{\alpha}}\right)$, gradient term $\widetilde{g} \equiv \left(g_{33} - \frac{f_{33}^2}{c_{11}}\right)$ and total voltage $V = V_0 + V_b$. Note that conventional Debye screening radius is $R_d^2 = \frac{k_B T}{2N_{d0}^+ e^2}\left(\varepsilon_0\varepsilon_b + \frac{1}{\alpha}\right)$. Estimations for STO parameters gives $\sqrt{\kappa\widetilde{g}} \sim 0.4$ nm, $\widetilde{R}_d \leq 10$ nm and $\widetilde{g}/\widetilde{\alpha} \sim 1$ nm for zero voltage.

However, the difference between $\widetilde{\alpha}$ and $\alpha$ depends on the applied voltage, since average polarization $\langle P\rangle$ depends on the voltage and film thickness via the cubic equation $\left(\alpha + \widetilde{\alpha}_{11}\langle P\rangle^2\right)\langle P\rangle = U/h$, which solution has the form:

$$\langle P\rangle = p(h,U)\left(1 - \frac{\alpha}{3\widetilde{\alpha}_{11}p(h,U)^2}\right) \approx \begin{cases} \dfrac{U}{\alpha h}, & \dfrac{U}{h} \ll 2\widetilde{\alpha}_{11}\left(\dfrac{\alpha}{3\widetilde{\alpha}_{11}}\right)^{3/2} \\ \left(\dfrac{U}{\widetilde{\alpha}_{11}h}\right)^{1/3}, & \dfrac{U}{h} \gg 2\widetilde{\alpha}_{11}\left(\dfrac{\alpha}{3\widetilde{\alpha}_{11}}\right)^{3/2} \end{cases}. \quad (A.8)$$

Where $p(h,U) = \left(\dfrac{U}{2\widetilde{\alpha}_{11}h} + \sqrt{\left(\dfrac{\alpha}{3\widetilde{\alpha}_{11}}\right)^3 + \left(\dfrac{U}{2\widetilde{\alpha}_{11}h}\right)^2}\right)^{1/3}$. Two limiting cases of Eq.(A.8) corresponds to low and high voltages respectively.

In decoupling approximation (i.e. assuming $P_i \approx -(\alpha^{-1})\partial\varphi/\partial x_i$ as described in subsection 2.4) the surface displacement has the form:

$$u_3^{decoupled}(0) = \left(\left(\frac{(\Xi - \beta)eU}{c_{11}k_B T}n_0 R_d - \frac{f_{1111}}{\alpha c_{11}}\frac{U}{R_d}\right)\tanh\left(\frac{h}{2R_d}\right) + \frac{q_{1111}^{MT}}{\alpha^2 c_{11}}\frac{U^2}{4R_d}\sinh^{-2}\left(\frac{h}{R_d}\right)\left(\frac{2h}{R_d} + \sinh\left(\frac{2h}{R_d}\right)\right)\right)$$
(A.9)

Note, that coupled equation Eq.(A.6) transfers into decoupled Eq.(A.9) in the limiting case $\tilde{g} \to 0$ and $\tilde{\alpha}_{11} \to 0$.

## Appendix B. Displacement for the well-localized probe potential

For cubic symmetry the convolution in the right-hand-side of Eq.(12a)

$$u_i(x_1, x_2, z) = -\iiint_{0<\xi_3<h} \frac{\partial G^S_{ij}(x_1-\xi_1, x_2-\xi_2, z, \xi_3)}{\partial \xi_m} \left( \begin{array}{c} \Xi_{mj}(n(\xi)-n_0) + \beta_{mj}(N^+_d(\xi)-N^+_{d0}) \\ + \frac{q^{MT}_{ijkl}}{\alpha^2}\frac{\partial \varphi}{\partial \xi_k}\frac{\partial \varphi}{\partial \xi_l} - \frac{f_{mjkl}}{\alpha}\frac{\partial^2 \varphi}{\partial \xi_k \partial \xi_l} \end{array} \right) d^3\xi \quad (B.1)$$

has explicit form consisting of three contributions for cubic symmetry:

(a) *concentration-strain* contribution

$$u_{CS} = -\iiint_{0<\xi_3<h} \frac{\partial G^S_{ij}(x_1-\xi_1, x_2-\xi_2, z, \xi_3)}{\partial \xi_m} \left( \Xi_{mj}(n(\xi)-n_e) + \beta_{mj}(N^+_d(\xi)-N^+_{de}) \right) d^3\xi$$

$$= -\iiint_{0<\xi_3<h} \frac{\partial G^S_{ij}(x_1-\xi_1, x_2-\xi_2, z, \xi_3)}{\partial \xi_j} \left( \Xi(n(\xi)-n_e) + \beta(N^+_d(\xi)-N^+_{de}) \right) d^3\xi$$

(B.2a)

(since $\Xi_{ij} = \Xi\delta_{ij}$ and $\beta_{ij} = \beta\delta_{ij}$ for cubic symmetry);

(b) *electrostriction* contribution

$$u_{ES} = -\iiint_{0<\xi_3<h} \frac{\partial G^S_{ij}(x_1-\xi_1, x_2-\xi_2, z, \xi_3)}{\partial \xi_m} q_{mjkl} P_k P_l \, d^3\xi$$

$$\equiv -\iiint_{0<\xi_3<h} \left( q_{1122}\left(\frac{\partial G^S_{ij}}{\partial \xi_j}\right)(P_k P_k) + (q_{1111}-q_{1122})\frac{\partial G^S_{ij}}{\partial \xi_j}P_j^2 + 2q_{1212}\sum_{j \neq k}\frac{\partial G^S_{ij}}{\partial \xi_k}P_k P_j \right) d^3\xi \quad (B.2b)$$

$$\xrightarrow{DA} -\iiint_{0<\xi_3<h} \left( \frac{q_{1122}}{\alpha^2}\frac{\partial G^S_{ij}}{\partial \xi_j}\frac{\partial \varphi}{\partial \xi_k}\frac{\partial \varphi}{\partial \xi_k} + \frac{q_{1111}-q_{1122}}{\alpha^2}\sum_j\frac{\partial G^S_{ij}}{\partial \xi_j}\left(\frac{\partial \varphi}{\partial \xi_j}\right)^2 + \frac{2q_{1212}}{\alpha^2}\sum_{j \neq k}\frac{\partial G^S_{ij}}{\partial \xi_k}\frac{\partial \varphi}{\partial \xi_k}\frac{\partial \varphi}{\partial \xi_j} \right) d^3\xi$$

can be presented as the "hydrostatic" part and shear part. DA is the abbreviation of the decoupling approximation.

(c) *Maxwell stress* contribution:

$$u_{MT} = -\iiint_{0<\xi_3<h} \frac{\partial G^S_{ij}(x_1-\xi_1, x_2-\xi_2, z, \xi_3)}{\partial \xi_m} \left( \delta_{mk}\delta_{jl} - \frac{\delta_{mj}\delta_{kl}}{2} \right)(P_k + \varepsilon_0 E_k) E_l \, d^3\xi$$

$$\approx \iiint_{0<\xi_3<h} d^3\xi \left( \frac{\partial G^S_{il}}{\partial \xi_k}P_k E_l - \left(\frac{\partial G^S_{ij}}{\partial \xi_j}\right)\frac{P_k E_k}{2} \right)$$

(B.2c)

Approximate equality in Eq.(B.2c) is valid for (quantum) paraelectrics, since their dielectric permittivity $\varepsilon \gg 1$.

(d) ***flexoelectric*** contribution

$$u_{FLEXO} = -\iiint_{0<\xi_3<h} \frac{\partial G_{ij}^S(x_1-\xi_1, x_2-\xi_2, z, \xi_3)}{\partial \xi_m} f_{mjkl} \frac{\partial P_l}{\partial \xi_k} d^3\xi$$

$$\equiv -\iiint_{0<\xi_3<h} \left( f_{1122} \left(\frac{\partial G_{ij}^S}{\partial \xi_j}\right)\left(\frac{\partial P_k}{\partial \xi_k}\right) + (f_{1111}-f_{1122})\frac{\partial G_{ij}^S}{\partial \xi_j}\frac{\partial P_j}{\partial \xi_j} + f_{1212}\sum_{j\neq k}\frac{\partial G_{ij}^S}{\partial \xi_k}\left(\frac{\partial P_j}{\partial \xi_k}+\frac{\partial P_k}{\partial \xi_j}\right)\right) d^3\xi \quad \text{(B.2d)}$$

$$\xrightarrow{DA} -\iiint_{0<\xi_3<h} \left( \begin{array}{l} \frac{f_{1111}+f_{1122}}{\alpha}\frac{1}{2}\frac{\partial G_{ij}^S}{\partial \xi_j}\Delta\varphi(\xi) + \\ \frac{f_{1111}-f_{1122}}{\alpha}\left(\sum_j \frac{\partial G_{ij}^S}{\partial \xi_j}\frac{\partial^2 \varphi}{\partial \xi_j^2} + \sum_{j\neq k}\frac{\partial G_{ij}^S}{\partial \xi_k}\frac{\partial^2 \varphi}{\partial \xi_k \partial \xi_j} - \frac{1}{2}\frac{\partial G_{ij}^S}{\partial \xi_j}\Delta\varphi(\xi)\right) \end{array} \right) d^3\xi$$

where we used that $2f_{1212} \equiv f_{1111} - f_{1122}$ for the isotropic media. Here the first term corresponds to "hydrostatic" part, while the latter is "shear" strains contribution.

After x,y-Fourier transformation

$$u_3(x_1, x_2, 0) = \int_{-\infty}^{\infty}\frac{dk_2}{2\pi}\int_{-\infty}^{\infty} dk_1 \exp(-ik_1 x_1 - ik_2 x_2)\tilde{u}_3(k_1, k_2, \xi_3) \quad \text{(B.3)}$$

and using Percival theorem we rewrite Eq.(B.1) as

$$\tilde{u}_i(k_1, k_2, x_3) = -\int_0^h d\xi_3 \tilde{G}_{ij,m}^S(k_1, k_2, x_3, \xi_3)\left( \begin{array}{l} \Xi_{mj}\delta\tilde{n}(k_1,k_2,\xi_3) + \beta_{mj}\delta\tilde{N}_d^+(k_1,k_2,\xi_3) \\ -\frac{f_{mjnl}}{\alpha}\tilde{\varphi}_{,nl}(k_1,k_2,\xi_3) + \frac{q_{ijkl}^{MT}}{\alpha^2}\frac{\partial\varphi}{\partial\xi_k}\frac{\partial\varphi}{\partial\xi_l} \end{array} \right). \quad \text{(B.4)}$$

Here

$$\tilde{G}_{ij,l}(k_1,k_2,\xi) \equiv \begin{cases} ik_l \tilde{G}_{ij}(k_1,k_2,\xi), & l=1,2 \\ \frac{\partial}{\partial\xi}\tilde{G}_{ij}(k_1,k_2,\xi), & l=3 \end{cases} \quad \text{(B.5a)}$$

$$\tilde{\varphi}_{,l}(k_1,k_2,\xi) \equiv \begin{cases} -ik_l \tilde{\varphi}(k_1,k_2,\xi), & l=1,2 \\ \frac{\partial}{\partial\xi}\tilde{\varphi}(k_1,k_2,\xi), & l=3 \end{cases} \quad \text{(B.5b)}$$

$$\tilde{G}_{3j}(k_1,k_2,\xi) = -\frac{1+\nu}{2\pi Y}\cdot\frac{ik_j \exp(-\xi k)}{k^2}(\xi k - (1-2\nu)), \quad j=1,2 \quad \text{(B.5c)}$$

$$\widetilde{G}_{33}(k_1,k_2,\xi) = \frac{1+\nu}{2\pi Y}\frac{\exp(-\xi k)}{k}(2(1-\nu)+\xi k). \tag{B.5d}$$

Here $k^2 = k_1^2 + k_2^2$,

Vertical displacement components

$$u_{CD}(\rho,0) = \int_0^\infty dk J_0(k\rho)k \int_0^\infty dz \exp(-kz)(2+2\nu)\left(\frac{\beta N_{d0}^+ e}{k_B T} - \frac{\Xi n_0 e}{k_B T}\right)$$
$$\times \widetilde{V}_0(k)\frac{\exp(-K(k)z) - \exp(-K(k)(2h-z))}{1 - \exp(-2K(k)h)} \tag{B.6}$$
$$= (2+2\nu)\left(\frac{\beta N_{d0}^+ e}{k_B T} - \frac{\Xi n_0 e}{k_B T}\right)\int_0^\infty dk J_0(k\rho)\frac{k\cdot \widetilde{V}_0(k)}{K(k)+k}$$

where we used the linear approximation $N_d^+ \approx N_{d0}^+\left(1+\frac{e\varphi}{k_B T}\right)$ and $n_0 \approx n_0\left(1-\frac{e\varphi}{k_B T}\right)$,

$$u_{FLEXO}(\rho,0) = \int_0^\infty \left(\frac{f_{1122}}{\alpha}\frac{\partial G_{ij}^S}{\partial \xi_j}\Delta\varphi(\xi) + \frac{f_{1111}-f_{1122}}{\alpha}\left(\sum_j \frac{\partial G_{ij}^S}{\partial \xi_j}\frac{\partial^2 \varphi}{\partial \xi_j^2} + \sum_{j\neq k}\frac{\partial G_{ij}^S}{\partial \xi_k}\frac{\partial^2 \varphi}{\partial \xi_k \partial \xi_j}\right)\right)d\xi_3 =$$
$$= \left(\begin{array}{l}-\dfrac{f_{1122}}{\alpha}\dfrac{(1+\nu)(1-2\nu)}{\pi Y(k+K(k))R_d^2} - \\ \dfrac{f_{1111}-f_{1122}}{\alpha}\left(\dfrac{(1+\nu)((1-\nu)2k+(1-2\nu)K(k))}{2\pi Y(k+K(k))^2 R_d^2} - \dfrac{(1+\nu)k^2}{\pi Y(k+K(k))^3 R_d^2}\right)\end{array}\right)\widetilde{V}_0(k) = \tag{B.7}$$
$$= \left(\begin{array}{l}-\dfrac{f_{1122}}{\alpha}\dfrac{(1+\nu)(1-2\nu)}{\pi Y(k+K(k))R_d^2} - \\ \dfrac{f_{1111}-f_{1122}}{\alpha}\left(\dfrac{(1+\nu)(1-2\nu)}{2\pi Y(k+K(k))R_d^2} + \dfrac{(1+\nu)k}{2\pi Y(k+K(k))^4 R_d^4}\right)\end{array}\right)\widetilde{V}_0(k)$$

where we used that $2f_{1212} \equiv f_{1111} - f_{1122}$ for the isotropic media, $K(k) = \sqrt{k^2 + R_d^{-2}}$

For the electrostriction contribution we used the approximation $\widetilde{\varphi}_V^2(k,z) \sim \dfrac{2}{R_0^2}\widetilde{V}_0\left(\dfrac{k}{2}\right)\widetilde{V}_0\left(\dfrac{k}{2}\right)\exp\left(-2K\left(\dfrac{k}{2}\right)z\right)$ that tends to the Dirac delta-function at $R_0 \to \infty$.

## Appendix C

Integration of Eq.(19b) with Gaussian potential (20) leads to

$$u_{CS}(0) = \frac{2(1+\nu)(1-2\nu)}{Y}(\Xi-\beta)N_{d0}^{+}\frac{V_0 e}{k_B T}R_d\left(1-\sqrt{\frac{\pi}{2}}\frac{R_d}{R_0}\left(1-\exp\left(\frac{R_0^2}{2R_d^2}\right)\mathrm{erf}\left(\frac{R_0}{\sqrt{2}R_d}\right)\right)\right) \quad \text{(C.1a)}$$

Integration of Eq.(19c) with Gaussian potential (20) leads to

$$u_{FLEXO}(0) = -\frac{(1+\nu)}{Y}\frac{V_0}{R_d}\left((1-2\nu)\frac{f_{1111}+f_{1122}}{\alpha}\left(1-\sqrt{\frac{\pi}{2}}\frac{R_d}{R_0}\left(1-\exp\left(\frac{R_0^2}{2R_d^2}\right)\mathrm{erf}\left(\frac{R_0}{\sqrt{2}R_d}\right)\right)\right)\right) -$$

$$-\frac{(1+\nu)}{Y}\frac{f_{1111}-f_{1122}}{\alpha}\frac{V_0}{R_0^5}\left(\begin{array}{l} 4\dfrac{\sqrt{\pi}}{\sqrt{2}}\left(30R_d^4-9R_d^2 R_0^2+R_0^4\right)\left(1-\exp\left(\dfrac{R_0^2}{2R_d^2}\right)\mathrm{erf}\left(\dfrac{R_0}{\sqrt{2}R_d}\right)\right)+\\ +\dfrac{\sqrt{\pi}}{\sqrt{2}}3R_0^2\left(20R_d^2-R_0^2\right)-4R_d R_0\left(30R_d^2+R_0^2\right)\end{array}\right)$$

(C.1b)

We derived Pade approximations

$$1-\sqrt{\frac{\pi}{2}}\frac{R_d}{R_0}\left(1-\exp\left(\frac{R_0^2}{2R_d^2}\right)\mathrm{erf}\left(\frac{R_0}{\sqrt{2}R_d}\right)\right) \approx \frac{R_0}{\sqrt{\dfrac{8}{\pi}R_d^2+\sqrt{2\pi}R_d R_0+R_0^2}} \quad \text{(C.2a)}$$

$$\left(\begin{array}{l} 4\dfrac{\sqrt{\pi}}{\sqrt{2}}\left(\dfrac{30R_d^4-9R_d^2 R_0^2+R_0^4}{R_0^5}\right)\left(1-\exp\left(\dfrac{R_0^2}{2R_d^2}\right)\mathrm{erf}\left(\dfrac{R_0}{\sqrt{2}R_d}\right)\right) \\ +\dfrac{\sqrt{\pi}}{\sqrt{2}}\dfrac{3\left(20R_d^2-R_0^2\right)}{R_0^3}-\dfrac{4R_d\left(30R_d^2+R_0^2\right)}{R_0^4}\end{array}\right) \approx$$

$$\approx \frac{R_0^2}{\dfrac{105}{8}R_d^3+\left(\dfrac{128}{\pi}-24\right)\sqrt{\dfrac{2}{\pi}}R_d^2 R_0+\dfrac{16}{\pi}R_d R_0^2+\sqrt{\dfrac{2}{\pi}}R_0^3} \approx \frac{R_0^2}{13R_d^3+13R_d^2 R_0+5R_d R_0^2+\sqrt{\dfrac{2}{\pi}}R_0^3}$$

(C.2b)

Integration of Eq.(19d) with Gaussian potential (20) leads to

(a) $\tilde{V}_0(k)V_b\left(q_{1111}^{MT}+q_{1122}^{MT}\right)$ term:

$$\int_0^\infty kR_0^2 \exp\left(-\frac{(kR_0)^2}{2}\right)\frac{2K(k)}{1+(k+K(k))R_d}dk =$$

$$= \frac{1}{R_d} - \sqrt{\frac{\pi}{2}}\frac{1}{R_0}\left(1-\exp\left(\frac{R_0^2}{2R_d^2}\right)\mathrm{erf}\left(\frac{R_0}{\sqrt{2}R_d}\right)\right) + \frac{\sqrt{\pi}}{R_d}\left(U\left(-\frac{1}{2}, 0, \frac{R_0^2}{2R_d^2}\right) - \frac{R_0}{R_d\sqrt{2}}\right) \approx \quad \text{(C.3a)}$$

$$\approx \frac{1}{R_d} - \sqrt{\frac{\pi}{8}}\frac{R_0}{(R_d+R_0)^2}$$

(b) $\tilde{V}_0(k)V_b\left(q_{1111}^{MT} - q_{1122}^{MT}\right)$ term:

$$\int_0^\infty kR_0^2 \exp\left(-\frac{(kR_0)^2}{2}\right)\frac{2kR_d(k+K(k))}{(1+(k+K(k))R_d)^2}dk =$$

$$= \frac{\sqrt{\pi}}{R_d}\left(U\left(-\frac{1}{2}, 0, \frac{R_0^2}{2R_d^2}\right) - \frac{R_0}{R_d\sqrt{2}}\right) \approx \frac{1}{\sqrt{R_d^2 + \sqrt{2\pi}R_d R_0 + \frac{8}{\pi}R_0^2}} \quad \text{(C.3b)}$$

$U(a, b, \xi)$ is the confluent hypergeometric function, which has the following integral representation $U(a, b, \xi) = \int_0^\infty \exp(-\xi\zeta)\zeta^{a-1}(1+\zeta)^{b-a-1}d\zeta \Big/ \int_0^\infty \exp(-\zeta)\zeta^{a-1}d\zeta$.

Next, using $\left(\tilde{V}_0(k/2)\right)^2/(2R_0^2) = V_0^2 R_0^2 \exp(-(kR_0)^2/4)/2$, similarly to flexoelectric contribution we obtained final approximate expression

$$u_{ES}(0) \approx -\frac{(1+\nu)}{Y}\left(\begin{array}{c}\frac{q_{1111}^{MT}+q_{1122}^{MT}}{\alpha^2}(1-2\nu)\left(V_0 V_b\left(\frac{1}{R_d}-\sqrt{\frac{\pi}{8}}\frac{R_0}{(R_d+R_0)^2}\right)+\frac{V_0^2 R_0}{2R_d\sqrt{\frac{4}{\pi}R_d^2+\sqrt{\pi}R_d R_0+R_0^2}}\right)\\ +\frac{q_{1111}^{MT}-q_{1122}^{MT}}{\alpha^2}\left(\frac{V_0 V_b}{\sqrt{R_d^2+\sqrt{2\pi}R_d R_0+\frac{8}{\pi}R_0^2}}+\frac{V_0^2 R_0^2}{13R_d^3+19R_d^2 R_0+10R_d R_0^2+\frac{4}{\sqrt{\pi}}R_0^3}\right)\end{array}\right)($$

C.4)

2 National Nanotechnology Initiative. 2006. http://www.nano.gov/.

3 P.I. Lazarev (Ed.) *Molecular Electronics*. Berlin: Springer (2004).

4 H. Masuhara and S. Kawata (Eds.) *Nanoplasmonics, Vol 2: From Fundamentals to Applications.* Elsevier, Amsterdam (2006).

5 S. Kawata, M. Ohtsu, and M. Irie (Eds.) *Nano-Optics*, Springer, Berlin (2002).

6 T.J. Chuang, P.M. Anderson, M.K. Wu, and S. Hsieh, *Nanomechanics of Materials and Structures*. Springer, Berlin (2006).

7 A.N. Cleland, *Foundations of Nanomechanics*, Springer (2002).

8 W.C. Oliver and G.M. Pharr, J. Mat. Res. **19**, 3 (2004).

9 B. Bhushan, *Nanotribology and Nanomechnics: An Introduction*, Springer (2008).

10 R.R. He and P.D. Yang, Nature Nano. **1**, 42 (2006).

11 S.V. Kalinin, B.J. Rodriguez, S. Jesse, B. Mirman, E. Karapetian, E.A. Eliseev, and A.N. Morozovska, Annu. Rev. Mat. Sci. **37**, 189 (2007).

12 W.G. Cady, *Piezoelectricity* (Vol. 1,2) Dover (1964).

13 M.E. Lines and A.M. Glass *Principles and Applications of Ferroelectric and Related Materials*. Clarendon Press, Oxford (1977).

14 L.L. Hench and J.K. West, *Principles of Electronic Ceramics*, Wiley, New York (1990).

15 N. Setter and E.L. Colla, *Ferroelectric Ceramics*, Birkhauser Verlag Basel (1993).

16 A.K. Tagantsev, Phys. Rev. **B 34**, 5883 (1986).

17. G. Catalan, L. J Sinnamon, J. M Gregg, J. Phys.: Condens. Matter **16,** 2253 (2004)

18. G. Catalan, B. Noheda, J. McAneney, L. J. Sinnamon, and J. M. Gregg, Phys. Rev B **72**, 020102 (2005).

19 R. Maranganti, N.D. Sharma, and P. Sharma, Phys. Rev. **B 74**, 014110 (2006).

20 S.V. Kalinin and V. Meunier, Phys. Rev. **B 77**, 033403 (2008).

21 M.S. Majdoub, P. Sharma, and T. Cagin, Phys. Rev. **B 77**, 125424 (2008).

22 A.K. Tagantsev, V. Meunier, and P. Sharma, MRS Bull. **34**, 643 (2009).

23 E.A. Eliseev, A.N. Morozovska, M.D. Glinchuk, and R. Blinc. Phys. Rev. B. **79**, 165433 (2009).

24 Y. T. Cheng, M. W. Verbrugge, J. Power Sources **190**, 453 (2009).

25 X. Zhang, A. M. Sastry, W. Shyy, J. Electrochem. Soc. **155**, A542 (2008).